\documentclass[12pt]{article}
\pdfoutput=1

\usepackage{amsmath,amssymb}
\usepackage[bookmarks=true]{hyperref}
\usepackage[nosort]{cite}
\usepackage[bulletsep]{collect}
\def\gfxon{\usepackage[final]{graphicx}}

\gfxon

\makeatletter
\let\old@startsection=\@startsection
\renewcommand{\@startsection}[6]{\old@startsection{#1}{#2}{#3}{#4}{#5}{#6\mathversion{bold}}}
\makeatother

\newcommand{\dpod}[1]{\partial_{#1}}

\newcommand{\lagr}{\mathcal{L}}

\makeatletter \@addtoreset{equation}{section} \makeatother

\makeatletter
\let\old@makecaption=\@makecaption
\def\@makecaption{\small\old@makecaption}
\makeatother

\newcommand{\ssN}{\mathcal{N}}

\newcommand{\me}{\mathrm{e}}

\newcommand{\Nt}{\widetilde N}

\makeatletter
\newcommand{\ellSN}{\mathop{\operator@font sn}\nolimits}
\newcommand{\ellCN}{\mathop{\operator@font cn}\nolimits}
\newcommand{\ellDN}{\mathop{\operator@font dn}\nolimits}
\newcommand{\ellAM}{\mathop{\operator@font am}\nolimits}
\newcommand{\ellK}{\mathop{\smash{\operator@font K}\vphantom{a}}\nolimits}
\newcommand{\ellE}{\mathop{\smash{\operator@font E}\vphantom{a}}\nolimits}
\makeatother

\ifx\genfrac\sdflkaj

\else

\fi
\newcommand{\sfrac}[2]{{\textstyle\frac{#1}{#2}}}
\newcommand{\half}{\sfrac{1}{2}}

\newcommand{\Integers}{\mathbb{Z}}

\newcommand{\Complex}{\mathbb{C}}

\newcommand{\CP}[1]{\mathbb{CP}^{#1}}

\def\[{\begin{equation}}
\def\]{\end{equation}}
\def\<{\begin{eqnarray}}
\def\>{\end{eqnarray}}

\def\beq{\begin{equation}}
\def\eeq{\end{equation}}
\def\beqn{\begin{eqnarray}}
\def\eeqn{\end{eqnarray}}

\newcommand{\nln}{\nonumber\\}
\newcommand{\earel}[1]{\mathrel{}&\hspace{-2\arraycolsep}#1\hspace{-2\arraycolsep}&\mathrel{}}

\newcommand{\tm}{\tilde{m}}
\newcommand{\ntwo}{${\mathcal N}=2\;$}
\newcommand{\ntt}{${\mathcal N}=(2,2)\,$}

\newcommand{\cell}{{\mathcal L}}

\newcommand{\pt}{\partial}

\newcommand{\eq}{\earel{=}}

\newcommand{\gsim}{\lower.7ex\hbox{$
\;\stackrel{\textstyle>}{\sim}\;$}}
\newcommand{\lsim}{\lower.7ex\hbox{$
\;\stackrel{\textstyle<}{\sim}\;$}}

\newcommand{\tN}{\widetilde{N}}

\newcommand{\hypref}[2]{\ifx\href\asklfhas #2\else\href{#1}{#2}\fi}

\newcommand{\NLSM}{NL\sigma M}
\newcommand{\GLSM}{GL\sigma M}

\def\slashed#1{\setbox0=\hbox{$#1$}             
   \dimen0=\wd0                                 
   \setbox1=\hbox{/} \dimen1=\wd1               
   \ifdim\dimen0>\dimen1                        
      \rlap{\hbox to \dimen0{\hfil/\hfil}}      
      #1                                        
   \else                                        
      \rlap{\hbox to \dimen1{\hfil$#1$\hfil}}   
      /                                         
   \fi}                                        %

\begin{document}

\begin{flushright}
{FTPI-MINN-11/17} \\
{UMN-TH-3007/11}\\
7/18/11
\end{flushright}
\vspace{0.2cm}

\renewcommand{\thefootnote}{\arabic{footnote}}
\begin{center}%
{\Large\textbf{\mathversion{bold}
Quantum Dynamics of Low-Energy Theory on Semilocal Non-Abelian Strings
}
\par}

\vspace{0.5cm}%

\textsc{P. Koroteev$^{1}$, M. Shifman$^{1,2}$, W. Vinci$^{1,2}$ and A. Yung$^{2,3}$  }

\vspace{3mm}

$^1$\textit{University of Minnesota, School of Physics and Astronomy\\%
Minneapolis, MN 55455, USA}
\\
\vspace{.1cm}
$^2$\textit{William I. Fine Theoretical Physics Institute, University of Minnesota, \\%
Minneapolis, MN 55455, USA}
\\
\vspace{.1cm}
$^3$\textit{Petersburg Nuclear Physics Institute, Gatchina, \\%
St. Petersburg 188300, Russia}

\vspace{3mm}

\texttt{koroteev,shifman,wvinci,ayung@physics.umn.edu} \\

\par\vspace{0.7cm}


\textbf{Abstract}\vspace{5mm}

\begin{minipage}{12.7cm}
Recently a low-energy effective theory on non-Abelian semilocal vortices in \ntwo SQCD with the
U$(N)$ gauge group and $N +\Nt$ quark flavors was obtained in field theory \cite{Shifman:2011xc}. The result 
is exact in a certain limit of large infrared cut-off. The resulting model
 was called the $zn$ model. We  study  quantum dynamics of the $zn$ model in some detail. 
 First we solve it at large $N$ in the leading order. Then we compare our results with those
 of Hanany and Tong \cite{Hanany:2004ea} (the HT model) who based their derivation on a certain type-IIA formalism, rather
 than on a field-theory construction. In the 't Hooft limit of infinite $N$ both model's predictions are identical. 
 At finite $N$ our calculations agree with the Hanany--Tong results only
 in the BPS sector. Beyond the BPS sector there is no agreement between the $zn$ and HT models. 
 Finally, we study perturbation theory of the $zn$ model from various standpoints. 
  \end{minipage}

\vspace{3mm}

\vspace*{\fill}

\end{center}

\newpage

\newpage

\tableofcontents

\newpage

\section{Introduction}
\label{Sec:Intro}
\setcounter{equation}{0}

Dorey and collaborators observed \cite{Dorey:1999zk,Dorey:1998yh} that the BPS spectrum of the twisted 
mass-deformed two-dimensional $\ssN = (2,2)$ $\CP{N-1}$  sigma model coincides with that of the 
four-dimensional $\ssN = 2$ SU$(N)$ supersymmetric quantum chromodynamics (SQCD) with $N$ massive flavors
(in a certain vacuum). This correspondence holds upon identification of the holomorphic parameters of the two theories,  e.g.  
the masses and the strong coupling scales. 
Similarities between sigma models in two dimensions and gauge theories in four dimensions have been 
discussed for a long time, 
since the discovery of asymptotic freedom and instantons in the O(3) sigma model \cite{Polyakov:1975yp, Polyakov:1975rr}.
  The observation  \cite{Dorey:1999zk,Dorey:1998yh} showed
that these  similarities go beyond the qualitative level in some supersymmetric theories. 
The deep reasons for this coincidence were revealed thanks to the discovery of the 
non-Abelian vortices in the color-flavor locked phase of supersymmetric QCD \cite{Hanany:2004ea,Auzzi:2003fs,Hanany:2003hp,Shifman:2004dr,Shifman:2007ce,Eto:2006pg,Tong:2005un,Tong:2008qd}. 
The two-dimensional $\CP{N-1}$ sigma model is nothing other than the low-energy  description of the non-Abelian string. Excitations of the non-Abelian string  correspond to  states of the bulk SQCD which are confined on the strings. In particular, BPS kinks of the
$\CP{N-1}$ model are confined monopoles from the bulk perspective \cite{Shifman:2004dr}. No surprise then that
the kink spectrum exactly coincides with the monopole spectrum.

The above results were naturally generalized to SU$(N)$ supersymmetric QCD with  $N +\tilde{N}$ flavors
(i.e. the number of flavors is larger than that of colors).
In this case one deals with the so-called {\em semilocal} 
\cite{Vachaspati:1991dz,Hindmarsh:1991jq,Hindmarsh:1992yy,Preskill:1992bf,Achucarro:1999it}
non-Abelian strings. Hanany and Tong suggested a world-sheet model for such strings \cite{Hanany:2004ea} (the HT model\,\footnote{The target space of the nonlinear sigma model obtained this way is now noncompact.
In mathematics it is mostly known as an $\mathcal O(-1)^{\Nt}$ fibration over $\CP{N-1}$. }) 
from type-IIA brane considerations. The Hanany--Tong
model can be easily formulated as the strong coupling limit of a U$(1)$ gauge theory with $N$ positively charged fields and $\Nt$ negatively charged fields under this U$(1)$. 

 While the Hanany--Tong model is exactly the theory considered by Dorey and collaborators, it is 
 {\em not} the genuine effective theory on the semilocal string world sheet. 
 The program of the field-theoretic honest-to-god
 derivation started with Refs.  \cite{Shifman:2006kd,Eto:2006uw,Eto:2007yv}.
Very recently a breakthrough 
was achieved in \cite{Shifman:2011xc} with the derivation of the ``exact'' effective theory on semilocal strings
valid in the limit
 $ \log L \to\infty$, where $L$ is an infrared cut-off assumed to be very large.

This exact nonlinear sigma model, to which we will refer to as the $zn$ model, was proven to 
have a different target-space metric  than the HT model (albeit the same topology). 

Our task is to explore dynamics of the $zn$ model per se and in comparison with the HT model. 
It is crucial to explicitly demonstrate that the $zn$ model  has the same BPS spectrum as four-dimensional SQCD,
as it was noted previously \cite{Dorey:1999zk,Dorey:1998yh} with regards to the  HT model.
We show that this is indeed the case. Moreover in the 't Hooft limit of infinite $N$
the solutions of both models are identical. However, at finite $N$ the $zn$ and HT models are different in the non-BPS sectors.
In particular, they have distinct perturbation theories. We analyze perturbation theory in the $zn$ model and explain in which sense
one can use here the notion of a single $\beta$ function.

We prove that the $\beta$ functions of the $zn$ model coincide with that of the HT model at one loop. Thanks to supersymmetry, this is enough to show the correspondence of the exact twisted Veneziano-Yankielowicz-type superpotentials which encode the 
BPS mass formula in terms of the central charges of each state. We conclude that the two models agree in the BPS sectors. 

The paper is organized as follows. First, in Sec.~\ref{Sec:HananyTongModel}  we introduce and compare two-dimensional sigma models which have recently been discussed in the literature in the context of semilocal strings in SQCD:  the $zn$ model \cite{Shifman:2011xc} and the Hanany-Tong \cite{Hanany:2003hp, Hanany:2004ea} model. In Sec.~\ref{Sec:LargeN} we study the large-$N$ solution, which we use in Sec.~\ref{Sec:Spectrumzn} to determine the spectrum of the theory.  We present an exact twisted superpotential which encodes the BPS spectrum at finite $N$ in Sec.~\ref{Sec:twisted}. Finally, in Sec.~\ref{Sec:GeometricFormulations} we study vacuum manifolds and perturbation theories of these models in the geometric formulation. We summarize and conclude in   Sec.~\ref{Sec:Consclusions}.

\section{World-Sheet Theory on Non-Abelian \\
Semi-Local Vortices: the \boldmath{$zn$} Model }
\label{Sec:HananyTongModel}
\setcounter{equation}{0}

Non-Abelian semilocal vortex strings (strings for short) are known to be supported by \ntwo SQCD
 with $N_{f}=N+\Nt$ massless flavors and the U$(N)$ gauge group \cite{Hanany:2003hp,Hanany:2004ea}
 provided one introduces a non-vanishing Fayet-Iliopoulos term 
 $\xi$.  Actually, the correct topological object to examine in connection
with the
semilocal strings is the second homotopy group of the vacuum manifold, which in the present case, is a Grassmannian  manifold (defined as follows):
\beq
\pi_{2}(\mathcal {M}_{\rm vac})=\pi_{2}\left({\rm Gr}_{N,\tilde N}\right)\equiv \pi_{2}\left(\frac{{\rm SU}(N+\tN)}{{\rm SU}(N)\times {\rm SU}(\tN)\times {\rm U}(1)}\right)=\mathbb{Z}\,.
\label{217}
\eeq 
The homotopy group above 
is the one lying behind the description of lumps in the associated nonlinear sigma-model,
 which arises as the low-energy limit of the \ntwo SQCD. 
 This is the main reason why semilocal strings are similar to lumps 
 \cite{Hindmarsh:1992yy,Eto:2006uw,Eto:2007yv}. Similarly to lumps, the 
 semilocal strings have power-law behaviors at large distances, and possess new {\em size} moduli  determining their characteristic thickness. Nevertheless, they still retain their nature of strings (flux tubes),   which is manifest when we send 
 the size moduli to zero. In this limit we recover just the ANO string, with its exponential behavior \cite{Vachaspati:1991dz}. The stringy nature is also justified by the existence of the following non-trivial homotopy group:
  \beq
 \pi_{1}({\rm U}(1)\times {\rm SU}(N)/ \mathbb{Z}_N )= \mathbb{Z}\,.
 \eeq
 
The moduli space of a single semilocal string is a non-compact 
 space of complex dimension $N+\Nt$ \cite{Hanany:2003hp,Shifman:2006kd,Eto:2007yv}. One can interpret $N-1$ zero modes as parameterizing orientational degrees of freedom
 of the non-Abelian string\footnote{The moduli space of a non-Abelian semilocal string contains indeed a subspace which corresponds to  $\CP{N-1}$ ,  the orientational moduli space of a traditional non-Abelian string.}, while further $\Nt$ modes parameterize the size(s) of the semilocal string. Finally, one last parameter 
 is due to translational modes; it is related to the position of the string center on the perpendicular plane. 
 Dynamically the latter moduli is decoupled from the rest. The corresponding dynamics is sterile.
 In the remainder of the paper it will be not mentioned. Then by the moduli space we will understand the 
 $(N+\Nt-1)$-dimensional manifold.

A crucial property of semilocal strings is
that, in deriving the world-sheet theory, one encounters an infrared divergence of the type
\beq
\log \frac{L}{|\rho|}\,,
\label{IRlog}
\eeq
regularized by an infrared (IR) cutoff $L$. Here $\rho$ is the typical size of a semilocal vortex. The above  logarithmic divergence is due to
long-range tails of the semilocal string which fall off as {\em powers} of the distance from the string axis 
(in the perpendicular plane) rather than exponentially. In the non-Abelian semilocal
strings both the size and orientational moduli become logarithmically 
non-normalizable \cite{Shifman:2006kd}.
A convenient and natural IR regularization, which maintains the
BPS nature of the solution\,\footnote{Alternatively, $L$ can   represent a finite length of the string, or a finite volume of the transverse space. } can be provided by a  mass difference
$\Delta m\neq 0$ of the (s)quark masses; then
  $L\sim 1/|\Delta m |$, so that (\ref{IRlog}) becomes
   \beq
\log \frac{1}{|\rho||\Delta m |}\,.
\label{IRlogp}
\eeq 

\subsection{The \boldmath{$zn$} model}
\label{tbznm}

 These large logarithms account for basically all difficulties in the previous treatments
 of the semilocal strings.   Such  divergent terms were calculated e.g. in Refs.   \cite{Shifman:2006kd,Eto:2007yv}.
 The situation was dramatically reversed  in \cite{Shifman:2011xc}. In this work the problem
 became  an advantage: {\em all} logarithmic terms were obtained from the bulk-theory description of the semilocal string.
 Then,  one can  derive an {\em exact}  world-sheet theory for the
 semilocal strings in the limit of (\ref{IRlog}) or (\ref{IRlogp}) tending to $\infty$. 
 The resulting model, which was called the $zn$ model,  is   $\mathcal N=(2,2)$ supersymmetric
 theory with 
  the following action\footnote{Here we write down only the bosonic part of the action;  we will include fermions in Sec.~\ref{Sec:LargeN}. } 
\begin{eqnarray}
S_{zn}
 &=& \int d^2 x\left\{ 
\frac1{4e^2}F^2_{kl} + \frac1{e^2}\,
\left|\pt_{k}\sigma\right|^2+ \frac{e^2}{2} \left(|n_{i}|^2 -r \right)^2
\right.
\nonumber\\[3mm]
&+&
\left|\pt_k(z_jn_i)\right|^2 + 
 \left|\nabla_{k} n_{i}\right|^2 +
\left.
|m_i-\widetilde m_{j}|^2 \,|z_j|^2|n_i|^2 
+\left|\sqrt{2}\sigma+m_i\right|^2 \left|n_{i}\right|^2 
\right\},
\nonumber\\[4mm]
&& 
i=1,...,N\,,\qquad j=1,...,\tN\,,\qquad \nabla_{k}=\pt_{k}-iA_{k}\,.
\label{Sgauge}
\end{eqnarray}
Here $n_i$ and $z_j$ are the orientational and size moduli fields, respectively,
$e^2$ and $r$ are the gauge coupling and the two-dimensional Fayet-Iliopoulos. In deriving the effective action above from the four-dimensional bulk theory one finds the crucial relationship between four and two dimensional couplings \cite{Hanany:2003hp,Shifman:2004dr}:
\beq
r=\frac{4 \pi}{g^{2}_{4\rm D}}\,.
\label{dtc}
\eeq
Finally, $m_i$ and $\widetilde m_{j}$ are twisted masses\footnote{These twisted masses are equal to the four-dimensional complex masses present in the bulk theory.}.
It is assumed that at the very end we take the limit $e\to\infty$.
In this limit the gauge field $A_{k}$  and its superpartners
become nondynamical, auxiliary \cite{Witten:1978bc,Hanany:1997vm} and can be integrated out
\beq
A_k=-\frac{i}{2r}(\bar{n}_i\pt_k n_i- n_i\pt_k \bar{n}_i), \qquad \sqrt{2}\sigma=- \frac{1}{r}\sum_{i}m_i\,|n_i|^2.
\label{eq:int1}
\eeq
Moreover, in this limit the term $\left(|n_{i}|^2 -r \right)^2$ in Eq.~(\ref{Sgauge}) implies the constraint\,\footnote{We stress that
this constraint is different from that in the Hanany--Tong model, see below.} 
\beq
\sum_{i}^{N}|n_{i}|^{2}=r\,.
\label{eq:int2}
\eeq
The fact that the number of degrees of freedom following from (\ref{Sgauge}) is correct,
namely,  $N+\Nt-1$, can be seen once we take into account  the $D$-term condition (\ref{eq:int2}) and, in addition,
gauge away a U$(1)$ phase.
The global symmetry of the world--sheet theory (\ref{Sgauge}) is the same as in that of the bulk theory,
\beq
{\rm SU}(N)\times {\rm SU}(\tN)\times {\rm U}(1)\,,
\label{globalsym}
\eeq
which is broken down to U(1)$^{N+\tilde{N}-1}$ by the (s)quark mass differences. 

\subsection{The HT model}

As was already mentioned, non-Abelian semilocal strings were previously studied within a  string theory approach based on D-branes by Hanany and Tong (see \cite{Hanany:1996ie, Hanany:2004ea} for the IIB setup and \cite{Hanany:2004ea} for the IIA setup). In the IIA picture a flux tube is represented by a D2-brane stretched between an NS5 and D4 branes. The effective theory on the world-sheet of the D2-brane, is then given by the strong-coupling limit ($e\to\infty$) of a  two-dimensional U$(1)$ gauge theory with $N$ positive and $\tN$ negatively  charged matter superfields. In components it reads
\begin{eqnarray}
S_{\rm HT} &=& \int d^2 x \left\{
\frac1{4e^2}F^2_{kl} + \frac1{e^2}\,
|\pt_k\sigma|^2+
 \frac{e^2}{2} \left(|n_i^{w}|^2-|z_j^w|^2 -r\right)^2
\right.
\nonumber\\[3mm]
&+&\left.
 |\nabla_{k} n_i^{w}|^2 +|\widetilde{\nabla}_{k} z_j^w|^2+
\left|\sqrt{2}\sigma+m_i\right|^2 \left|n_i^{w}\right|^2 
+ \left|\sqrt{2}\sigma+\widetilde m_{j}\right|^2\left|z_j^w\right|^2\right\},
\nonumber\\[3mm]
&& 
i=1,...,N,\qquad j=1,...,\tN\,, 
\nonumber\\[4mm]
&& 
{\nabla}_k=\pt_k-iA_k\,, \qquad \widetilde{\nabla}_k=\pt_k+iA_k\,.
\label{wcp}
\end{eqnarray}
With respect to the U(1) gauge field $A_k$
 the fields $n_i^{w}$ and $z_i^w$ have
charges  +1 and $-1$, respectively. We endow these fields with a superscript ``$w$'' (weighted) to distinguish them from 
the $n_i$ and $z_j$ fields which appear in the $zn$ model, see (\ref{Sgauge}). 
If only  charge $+1$ fields were present, in the limit  $e\to\infty$ we would get a conventional twisted-mass deformed
$\CP{N-1}$  model. The Hanany-Tong model can be obtained by the dimensional reduction (from 4D to 2D)
of the supersymmetric quantum electrodynamics with $N$ charge 1 and $\tN$ charge $-1$ chiral superfields.

\section{ \boldmath{$\beta$} function}
\label{Sec:beta}
\setcounter{equation}{0}

Let us calculate the one-loop renormalization of the coupling constant $r$ in the $zn$ model (\ref{Sgauge}).
To this end we can limit ourselves to the massless case $m_i=\tm_{j}=0$. Then the action (\ref{Sgauge}) 
can be rewritten as 
\beq
S_{\rm zn} = \int d^2 x\left\{ \left|\pt_k(z^jn^i)\right|^2 + 
 \left|\nabla_{k} n^{i}\right|^2 
+ iD\left(|n_{i}|^2 -r_0 \right)
\right\},
\label{znm0}
\eeq
where $r_0$ is a bare coupling constant and the limit $e\to\infty$ is taken. Integration over 
the auxiliary
field $D$ ensures the condition (\ref{eq:int2}), while the gauge field is given by
\beq
A_k=-\frac{i}{2|n|^2}\,(\bar{n}_i\pt_k n^i -n^i\pt_k \bar{n}_i).
\label{Ak}
\eeq
Next, we  rearrange the kinetic term by decomposing
\beq
\pt_k(z^jn^i)=z^j \nabla_{k} n^{i} +n^i \widetilde{\nabla}_{k} z^{j}\,.
\eeq
As a result, the  action (\ref{znm0})  takes the form
\begin{eqnarray}
S_{\rm zn} &=& \int d^2 x \left\{\rule{0mm}{6mm}
 \left|\nabla_{k} n'^{\,i}\right|^2 +\left|\widetilde{\nabla}_{k} z'^{\,j}\right|^2 + iD'\left(\left|n'^{\,i}\right|^2 -
 \left|z'^{\,j}\right|^2-r_0 \right)
 \right.
\nonumber\\[3mm]
&+&
\frac1{\left| n'\right|^2}\left(z'\nabla_{k}\bar{z}'\right)\left(\bar{n}'\nabla_{k}n'\right)
+\frac1{|n'|^2}\left(\bar{z}'\widetilde{\nabla}_{k}z'\right)\left(n'\widetilde{\nabla}_{k}\bar{n}'\right)
\nonumber\\[3mm]
&-&
\left.
\frac1{2|n'|^2}\left(\pt_k |n'|^2\right)\left(\pt_k |z'|^2\right)-\frac1{4|n'|^2}\, \left(\pt_k |z'|^2\right)^2
\right\},
\label{htlike}
\end{eqnarray}
where we introduced new variables
\beq
n'^i=\sqrt{1+|z|^2}\,n^i,\qquad z'^j=\sqrt{r_0}\,z^j,\qquad D'=\frac1{1+|z|^2}\,D\,,
\label{newfields}
\eeq
and the indices $i,j$ are contracted in the brackets, e.g. $\left(z'\nabla_{k}\bar{z}'\right) \equiv \left(z'^{\,j}\nabla_{k}\bar{z}'_j\right)$. In passing from (\ref{znm0}) to  (\ref{htlike}) we used the constraint $|n|^2 = r_0$.
Solving the equations of motion for the gauge potential $A_k$ in (\ref{htlike})  we find that it is still given by
Eq.~(\ref{Ak}), as it should, of course.

A disadvantage of formulation (\ref{htlike}) in terms of $n'$ and $z'$ is rather obvious: change of variables (\ref{newfields})
is not  holomorphic and, therefore, the metric of the target manifold in (\ref{htlike})   does not explicitly look as
a  metric of a K\"ahler manifold. Certainly, we know that the model (\ref{Sgauge})
is \ntt supersymmetric and has a K\"ahler target-space metric in terms of the original fields $n$, $z$. 

The action (\ref{htlike}) reveals a similarity between the $zn$ model and the HT model (\ref{wcp}). 
In particular, the  first line in (\ref{htlike})
is identical to the massless limit of the HT model (\ref{wcp}) at $e\to\infty$.
Moreover, all terms in the second and third lines in (\ref{htlike}) do not contribute at  one-loop. Therefore,
we conclude that the one-loop renormalization of the coupling constant $r$ is identical in the $zn$ and
HT models.

More explicitly, to calculate the one-loop renormalization of  $r$ we represent the fields $n'$ and $z'$
in (\ref{htlike}) as  sums of classical background fields plus quantum fluctuations,
\beq
n'^{\,i}=n_0^i+\delta n^i,\qquad z'^j=z_0^j+\delta z^j.
\label{qufluct}
\eeq
The renormalization of $r$ can be calculated as that of the linear  in $D'$ term in (\ref{htlike}).
Let us  write the third term in the first line in (\ref{htlike}) as 
\beq
iD'\left(|n_0^{i}|^2 -|z_0^j|^2 +|\delta n^{i}|^2 -|\delta z^j|^2-r_0 \right).
\eeq
It contributes to the one-loop renormalized coupling $r$
\beq
r_{\rm ren}=r_0-\langle\, |\delta n^{i}|^2\, \rangle +\langle\, |\delta z^j|^2\,\rangle \,,
\label{rrenormalization}
\eeq
where $\langle ... \rangle $ stands for vacuum averaging.

Calculating the one-loop tadpole contributions  here using canonical propagators
of $n'$ and $z'$ fields defined by the first line in (\ref{htlike}) we get 
\beq
r_{\rm ren}(\mu)=r_0 -(N-\Nt)\int \frac{d^2k}{(2\pi)^2}\frac{1}{k^2}\,= r_0-\frac{N-\tN}{2\pi}\,\log{\frac{M}{\mu}},
\label{rrenM}
\eeq
where $M$ is the ultraviolet cutoff, while $\mu$ is the infrared normalization point. The terms 
proportional to $N$  and $\tN$ arise due to loops of $n'$ and $z'$ fields, respectively.
Introducing the dynamical scale of the theory $\Lambda$,
\beq
\Lambda \equiv M \exp\left(- \frac{2\pi \, r_0}{N-\tilde N}\right) ,
\eeq
 we rewrite (\ref{rrenM}) as 
\beq
r_{\rm ren}(\mu)=\frac{N-\tN}{2\pi}\,\log{\frac{\mu}{\Lambda}}.
\label{rren}
\eeq
The $zn$ model is asymptotically free at $N>\tN$ (which is assumed throughout the paper). The 
one-loop renormalization of its coupling
constant is identical to that of the HT model calculated in \cite{Hanany:1997vm}.

The coupling constant $r$  can be complexified by adding a $\theta$ term in the theory. The target space in the model at hand 
is K\"ahlerian but non-Einstein\footnote{For an Einstein manifold, the Ricci tensor is proportional to the metric tensor.}. Therefore, $r$ does not completely specifies the one-loop renormalization group (RG) flow of this 
theory. We will discuss this question in more detail later. Here let us make a statement using
the HT model as an example (a similar statement can be formulated for the $zn$ model too).
Let us keep the coupling constant $e$ large but finite. Then we have two large parameters of mass
dimension one: the ultraviolet cutoff $M$ and $e$. The normalization point $\mu$ is supposed to be $\ll M$.
If $\mu\gg e$, then the effective action must be holomorphic in the complexified coupling $r$, implying
that higher loops cannot contribute to the $\beta$ function in this domain. The one-loop renormalization (\ref{rren})
 is actually exact both, in the $zn$ and HT models for such values of $\mu$. The holomorphicity is lost, 
 generally speaking, when we evolve $\mu$ below $e$, due to emergence of additional structures in the effective Lagrangian, see
 Sec.~\ref{Sec:GeometricFormulations}. In this domain the RG flow ceases to be one-loop. However, in the large-$N$ limit, in the leading in $N$ order, 
 the one-loop nature is preserved.

\section{Exact Effective Twisted Superpotentials}
\label{Sec:twisted}
\setcounter{equation}{0}

The one-loop calculation performed in the previous section can be enhanced by supersymmetry to give  exact results, as shown in Refs. \cite{Witten:1993yc,Dorey:1998yh} in both the regimes $e\ll \mu$   and $e\gg\mu$. To see this, first we recall that the renormalized Fayet-Iliopoulos in two-dimensional $\mathcal N=(2,2)$ must be written in  terms of a complex twisted superpotential $\widetilde W$ of Veneziano--Yankilowicz type \cite{Cecotti:1992rm,Witten:1993yc},
as dictated by supersymmetry:\beq
r_{\rm eff}=-\widetilde W_{\rm eff}'(\sigma)\,.
\eeq
Using the result of the previous section we can write down the following effective twisted superpotential for the $zn$ model in the case of the vanishing masses. 
\beq
\label{eq:Weffzeromass}
\widetilde{W}_{\rm eff} =  -\frac{N-\Nt}{2\pi}\sqrt2\sigma\left(\log\frac{\sqrt2\sigma}{\Lambda}-1\right)\,.
\eeq
The one-loop expression above is exact, thanks to holomorphicity,  in the regime $e\ll \mu$. Nevertheless, there are two important observations which makes the potential above a crucial tool for extracting exact results from the theory at all values of the coupling $e$. First notice that the twisted superpotential above does not depend on the gauge coupling $e$. This is due to the fact that only the couplings which can be promoted to twisted chiral superfields can appear in $\widetilde W$, and this is certainly not the case for $e$. The bottom line of this observations is that the all the information which can be extracted from this potential are actually exact, and also valid in the nonlinear sigma model limit when  $e\rightarrow\infty$. The second observation is that the difference of the values of the twisted superpotential $\widetilde W$  between two vacua gives  the central charges and thus the masses of the BPS states of the theory\footnote{Differences between different vacua give the masses of the solitonic states such as kinks. Since $\widetilde W$ is a multi-valued function, it makes sense to take differences between the values of  $\widetilde W$ taken between the same vacua but on different Riemann sheets. This will give masses of the perturbative spectrum.}
\beq
M_{\rm BPS} = |Z|=\Delta \widetilde W\,.
\eeq
Notice again that the mass formula written above is exact for all values of $e$. While it represents a perturbative calculation at small $e$, it encodes full non-perturbative corrections to the masses of all BPS states  in the regime  $e\rightarrow \infty$.

We wish to emphasize here that \eqref{eq:Weffzeromass} is exact only if applied to the BPS sector of the theory. Once we start looking at perturbations around the vacua given by minimization of the twisted superpotential, formula \eqref{eq:Weffzeromass}, or its massive generalization, is of no use. Still, when we treat the model in the large-$N$ approximation, the effective potential
\[\label{eq:VvsW}
V(\sigma) = \left|\widetilde{W}'_{\rm eff}\right|^2\,,
\]
give the correct spectrum of the theory. We will address both questions in the next section.

Finally let us note that twisted masses can be introduced in the 
theory by gauging each U(1) factor in the U(1)$^N_f$ group by its own
gauge field with non-zero $\sigma$-component (equal to associated mass)
\cite{Hanany:1997vm}. This leads to the following generalization of the effective twisted superpotential \eqref{eq:Weffzeromass} to the case of non-zero twisted masses:
%
\begin{eqnarray}
\label{eq:ExactTwistSP}
\widetilde{W}_{\rm eff} & =&   -\frac{1}{2\pi}\sum_{i=1}^N(\sqrt2\sigma+m_i)
\left(\log\frac{\sqrt2\sigma+m_i}{\Lambda}-1\right)+\nonumber \\
&+& \frac{1}{2\pi}\sum_{j=1}^{\Nt}(\sqrt2\sigma+\widetilde m_j)\left(\log\frac{\sqrt2\sigma+\widetilde m_j}{\Lambda}-1\right)\,.
\end{eqnarray}
Clearly this effective twisted superpotential identically coincides with the one for HT model \cite{Hanany:1997vm}.

This fact together with the matching of the kink spectrum obtained at the classical level in Ref. \cite{Shifman:2011xc}, leads us to claim the matching of the BPS spectra of the $zn$ and HT at both semiclassical and quantum levels. As a consequence, the BPS spectrum of the bulk theory coincides with the BPS spectrum of the true effective theory on semilocal vortices, as expected.

\section{Large-\boldmath{$N$} Solution of the \boldmath{$zn$} Model}
\label{Sec:LargeN}
\setcounter{equation}{0}

In this section we will study the $zn$ model at large $N$ along the lines of  Witten's analysis \cite{Witten:1978bc}.
Namely, we will consider the limit $N\to\infty$, $\tN\to\infty$, while the ratio of $\tN$ and $N$ is kept fixed.
The representations (\ref{wcp}) and (\ref{htlike}) suggest that to the leading order in $N$ the solutions of $zn$  and 
the HT models are the same. The reason for this is that all terms in the second and third lines in (\ref{htlike})  distinguishing the $zn$ model from the HT model give nonvanishing contributions only at a subleading 
order in $N$. Indeed,  they can show up in the potential for $\sigma$ only at the two-loop order and are not reducible to the $n'$ and $z'$ 
field tadpoles proportional to $N$ or $\tN$. Inspection of the SU$(N)$ and SU$(\Nt)$ index flow
readily reveals that these two- and higher-loop contributions are at most $O(N^0)$ in the large $N$-limit. 

Below we will calculate the effective action for the $zn$ model with twisted masses in the large-$N$ limit. The action of the $zn$ model (\ref{Sgauge}) in the gauged formulation, with the fermion fields taken into account,  is
\begin{eqnarray}
S_{\rm zn} &=& \int d^2 x\left\{ 
\frac1{4e^2}F^2_{kl} + \frac1{e^2}\,
\left|\pt_{k}\sigma\right|^2 +\frac1{2e^2}D^2 +\frac1{e^2}\,\bar{\lambda}_{R}\,i \pt_{L}\, \lambda_R
+\frac1{e^2}\,\bar{\lambda}_{L}\,i \pt_{R}\, \lambda_L
\right.
\nonumber\\[3mm]
&+&
\left|\pt_k(z^jn^i)\right|^2 + 
 \left|\nabla_{k} n^{i}\right|^2 
+|m_i-\widetilde m_{j}|^2 \,|z^j|^2|n^i|^2 
\nonumber\\[3mm]
&+&
\left|\sqrt{2}\sigma+m_i\right|^2 \left|n^{i}\right|^2 
+ iD \left(|n^{i}|^2 -r_0 \right)
\nonumber\\[3mm]
 &+&
 \bar{\xi}_{iR}\,i \nabla_{L}\, \xi^{i}_R
+ \bar{\xi}_{iL}\,i\nabla_{R}\, \xi^{i}_L
\nonumber\\[3mm]
 &+&
\Big[ i(\sqrt{2}\,\sigma+m_i)\,\bar{\xi}_{iR}\xi^i_L
+ i\sqrt{2}\,\bar{n}_i\,(\lambda_R\xi^i_L-
\lambda_L\xi^i_R) + {\rm H.c.} \Big]
\nonumber\\[3mm]
 &+&
(\bar{z}_j\bar{\xi}_{iL} + \bar{n}_i\bar{\chi}_{jL})\,i\pt_R (z^j\xi^i_L+n^i\chi^j_L)
+(\bar{z}_j\bar{\xi}_{iR} + \bar{n}_i\bar{\chi}_{jR})\,i\pt_L (z^j\xi^i_R+n^i\chi^j_R)
\nonumber\\[3mm]
 &+&
\Big[i(m_i-\widetilde{m}_j)\,\left(|z^j|^2\bar{\xi}_{iR}\xi^i_L + |n^i|^2\bar{\chi}_{jR}\chi^j_L
+\bar{\xi}_{iR}\chi^j_L \bar{z}_j n^i + \bar{\chi}_{jR}\xi^i_L \bar{n}_i z^j\right)
+ {\rm H.c.} \Big]
\nonumber\\[3mm]
 &+&
\left.
\bar{\chi}_{jR}\chi^j_R \bar{\xi}_{iL}\xi^i_L + \bar{\chi}_{jL}\chi^j_L \bar{\xi}_{iR}\xi^i_R
+\bar{\chi}_{jL}\chi^j_R \bar{\xi}_{iR}\xi^i_L + \bar{\chi}_{jR}\chi^j_L \bar{\xi}_{iL}\xi^i_R
\rule{0mm}{6mm}\right\},
\label{znferm}
\end{eqnarray}
where the fields $A_k$, $\sigma$, $D$ and $\lambda_{L,R}$ form the gauge supermultiplet, while $\xi^i$ and
$\chi^j$ are fermion superpartners of $n^i$ and $z^j$, respectively. Left and right derivatives are defined as
\beq
\nabla_L \equiv \nabla_{0}-i\nabla_3\,,\qquad \nabla_R \equiv \nabla_{0} + i\nabla_3\,.
\eeq

\subsection{Effective potential at large $N$}

Now we will integrate over the  $n^i$, $z^j$ and $\xi^i$,
$\chi^j$ fields and then minimize the resulting
effective action with respect to the  fields $\sigma $ and $D$  from the gauge multiplet.
This will be done in the saddle point approximation.
 The large-$N$ limit ensures that the corrections to the saddle point approximation (suppressed by $1/N$) are negligible. 
 
Technically, integrating out the  $n^i$, $z^j$ and $\xi^i$,
$\chi^j$ fields  in the saddle point
boils down to calculating a  set of one-loop graphs with the
$n^{i}$ and $z^j$  superfields propagating in loops.
As was mentioned, in this section we will obtain the effective potential of the theory as a function of $\sigma$ and $D$.
Minimization of this potential determines the vacuum structure of the theory. At this stage we can
 drop the gauge field $A_k$ and its fermion superpartners $\lambda_{L,R}$ in (\ref{znferm}) 
 because they have no vacuum values. If desirable, one can restore the $A_k$  dependence in the final result from
 gauge invariance, through  replacing partial derivatives by covariant.

Since the action (\ref{znferm}) is not quadratic in $n^i$, $z^j$ and $\xi^i$,
$\chi^j$ fields we do the integration in two steps. First, we integrate over $n^i$ and $\xi^i$.
It turns out that the resulting effective action will be quadratic in $z^j$ and $\chi^j$
and at the next stage we will be able to integrate out these fields too.

After rescaling the $n^i$ and $\xi^i$ fields  similar to that in (\ref{newfields}), namely,
\beq
n'^i=\sqrt{1+|z|^2}\,n^i,\qquad \xi'^i=\sqrt{1+|z|^2}\,\xi^i
\label{rescalen}
\eeq
integration over the bosonic fields gives the determinant
\beq
\prod_i^N \left[{\rm det}\, \left(-\pt_{k}^2 +\frac{iD}{1+|z|^2}
+M_{Bi}^2\right)\right]^{-1},
\label{detb}
\eeq
while the fermion integration gives
\beq
\prod_i^N  {\rm det}\, \left(-\pt_{k}^2 
+M_{Fi}^2\right),
\label{detf}
\eeq
where $M_B^2$ and $M_F^2$ are the following functions: 
\begin{eqnarray}
&&
M_{Bi}^2(\sigma , z^j,\chi^j)=\frac1{1+|z|^2}\left\{|\sqrt{2}\sigma+m_i\right|^2+|m_i-\widetilde m_{j}|^2 \,|z^j|^2
\nonumber\\[3mm]
 &+&
\left.
|\pt_k z^j|^2+\bar{\chi}_{jR}\,i \pt_{L}\, \chi^{j}_R
+ \bar{\chi}_{jL}\,i\pt_{R}\, \chi^{j}_L +i(m_i-\widetilde{m}_j)\bar{\chi}_{jR}\chi^j_L\rule{0mm}{5mm}\right\}
\label{Mb}
\end{eqnarray}
and 
\begin{eqnarray}
&&
M_{Fi}^2(\sigma , z^j,\chi^j)=\frac1{(1+|z|^2)^2}\left\{|\sqrt{2}\sigma+m_i\right|^2+\left|(m_i-\widetilde m_{j}) \,|z^j|^2\right|^2
\nonumber\\[3mm]
 &+&
(\sqrt{2}\bar{\sigma}+\bar{m}_i)(m_i-\widetilde m_{j})\,|z^j|^2 +(\sqrt{2}\sigma+m_i)
(\bar{m}_i-\bar{\widetilde{m}}_{j})\,|z^j|^2
\nonumber\\[3mm]
 &+&
\left.
i\Big[(\sqrt{2}\sigma+m_i)+(m_i-\widetilde m_{j})\,|z^j|^2\bar{\chi}_{jR}\chi^j_L+{\rm H.c.}\Big]
\right\}.
\label{Mf}
\end{eqnarray}
Calculating the determinants (\ref{detb}) and (\ref{detf}) gives the effective action
as a functional of the fields $\sigma$, $D$, $z^j$ and $\chi^j$,
\begin{eqnarray}
S_{\rm eff}(\sigma,D , z^j,\chi^j)
&=&
\int d^2 x\left\{\frac1{4\pi}\sum_{i=1}^{N}\left[\left(M^2_{Bi}+\frac{iD}{1+|z|^2}\right)\,
\log{\frac{M^2}{M^2_{Bi}+\frac{iD}{1+|z|^2}}} 
\right.\right.
\nonumber\\[4mm]
&+&
\left.\left.
\frac{iD}{1+|z|^2}+M^2_{Fi}\,\log{\frac{M^2}{M^2_{Fi}}} +M^2_{Bi}-M^2_{Fi}\right] -iDr_0\right\},\notag\\
\label{Szint}
\end{eqnarray}
where $M$ is the ultraviolet cut-off scale. 

Next, expand the action (\ref{Szint}) in powers of the fields $z^j$ and $\chi^j$. We see that
certain terms quadratic in these fields come with an infinitely large logarithmic $Z$-factors. 
This is a crucial point.
Say,
we get kinetic terms of the type
\beq
\Big\{|\pt_k z^j|^2+\bar{\chi}_{jR}\,i \pt_{L}\, \chi^{j}_R
+ \bar{\chi}_{jL}\,i\pt_{R}\, \chi^{j}_L\Big\}\,\log{\frac{M^2}{\mu^2}},
\eeq
where $\mu$ is some infrared scale determined by the value of $\sigma$ and twisted masses.
We absorb this infinite $Z$-factor redefining the fields $z^j$ and $\chi^j$ as
\beq
z'^j=\sqrt{\frac{N}{4\pi}\,\log{\frac{M^2}{\mu^2}}}\,\,z^j,\qquad
\chi'^j=\sqrt{\frac{N}{4\pi}\,\log{\frac{M^2}{\mu^2}}}\,\,\chi^j.
\label{newzchi}
\eeq
Now if we re-express the effective action (\ref{Szint}) in terms of new variables, we see that
higher powers of the $z'^j$ and $\chi'^j$ fields  are suppressed by powers of the large logarithm
and can be dropped. As a result, the effective action (\ref{Szint}) turns out to be quadratic in
 the $z'^j$ and $\chi'^j$ fields! Thus, we obtain
\begin{eqnarray}
&&
S_{\rm eff}(\sigma,D , z^j,\chi^j)
\nonumber\\[3mm]
&=&
\int d^2 x\left\{\frac1{4\pi}\sum_{i=1}^{N}\left[\left(|\sqrt{2}\sigma+m_i\right|^2+iD\right)\,
\log{\frac{M^2}{|\sqrt{2}\sigma+m_i|^2+iD}} +iD 
\right.
\nonumber\\[3mm]
 &-&
\left.
|\sqrt{2}\sigma+m_i|^2\,
\log{\frac{M^2}{|\sqrt{2}\sigma+m_i|^2}}\right]
+ |\pt_k z'^j|^2+\bar{\chi}'_{jR}\,i \pt_{L}\, \chi'^{j}_R
+ \bar{\chi}'_{jL}\,i\pt_{R}\, \chi'^{j}_L
\nonumber\\[3mm]
&-&
iD(r_0+|z'^j|^2)
\nonumber\\[4mm]
&+&
\left.
|\sqrt{2}\sigma+\widetilde m_{j}|^2\,|z'^j|^2-
\left[(\sqrt{2}\sigma+\widetilde m_{j})\bar{\chi}'_{jR}\chi'^j_L
+ {\rm H.c.}\right]\right\}.
\label{Sz}
\end{eqnarray}

Note, that the sign of the interaction term of $z'$ with $D$ (and $\chi_{L,R}'$ with $\sigma$) shows that the
$z'$ multiplet has charge $-1$, as was expected. One can restore the gauge field 
dependence in (\ref{Sz}) through the substitution
\beq
\pt_k\rightarrow \widetilde{\nabla}_k.
\eeq
Simultaneously, we will recover terms proportional to $\left(\bar{z}_j\pt_k z^j\right)$ and $\left(\bar{\chi}_{jL}\chi^j_L\right)$, $\left(\bar{\chi}_{jR}\chi^j_R\right)$. The $z'$ and $\chi_{L,R}'$-dependent part of the action (\ref{Sz})
is just the U(1) gauge theory of the $z'$ multiplet with charge $-1$ plus the FI $D$-term $r_0$.

Now, since the action (\ref{Sz}) is quadratic in the fields from the $z'$ multiplet we can integrate 
out $z'$ and $\chi_{L,R}'$. As a result, we arrive at  the effective potential as a function of
the fields $\sigma$ and $D$
\begin{eqnarray}
V_{\rm eff}(\sigma, D )&=&
 \frac{1}{4\pi}
\sum_{i=1}^{N}\left[\left(\left|\sqrt{2}\sigma+m_i\right|^2+iD\right)\,
\log{\frac{M^2}{|\sqrt{2}\sigma+m_i|^2+iD}} +iD 
\right.
\nonumber\\[3mm]
&-&
\left.
|\sqrt{2}\sigma+m_i|^2\,
\log{\frac{M^2}{|\sqrt{2}\sigma+m_i|^2}}
\right]
\nonumber\\[3mm]
&+&
\frac1{4\pi}\sum_{j=1}^{\tN}\left[\left(\left|\sqrt{2}\sigma+
\widetilde m_{j}\right|^2-iD\right)\,
\log{\frac{M^2}{|\sqrt{2}\sigma+\widetilde m_{j}|^2-iD}}
\right.
\nonumber\\[3mm]
 &-&
\left.
iD -|\sqrt{2}\sigma+\widetilde m_{j}|^2\,
\log{\frac{M^2}{|\sqrt{2}\sigma+\widetilde m_{j}|^2}}\right]
 -iD r_0\,.
\label{VeffM}
\end{eqnarray}
Using the $\beta$ function of the theory we can trade the
bare coupling $r_0$ here for the dynamical scale $\Lambda$,
by writing 
\beq
r_{0}=\frac{N-\tN}{2\pi}\,\log{\frac{M}{\Lambda}}.
\label{r0}
\eeq
Substituting this in (\ref{VeffM}) we see that the dependence on the ultraviolet cut-off scale $M$ cancels
out, and we get
\begin{eqnarray}
V_{\rm eff}(\sigma,D)
&=&
\frac1{4\pi}
\sum_{i=1}^{N}\left[-\left(|\sqrt{2}\sigma+m_i\right|^2+iD\right)\,
\log{\frac{|\sqrt{2}\sigma+m_i|^2+iD}{\Lambda^2}} +iD 
\nonumber\\[3mm]
 &+&
\left.
\left|\sqrt{2}\sigma+m_i\right|^2\,
\log{\frac{|\sqrt{2}\sigma+m_i|^2}{\Lambda^2}}\right]
\nonumber\\[3mm]
&+&
\frac1{4\pi}\sum_{j=1}^{\tN}\left[-\left(|\sqrt{2}\sigma+
\widetilde m_{j}\right|^2-iD\right)\,
\log{\frac{|\sqrt{2}\sigma+\widetilde m_{j}|^2-iD}{\Lambda^2}}
\nonumber\\[3mm]
 &-&
\left.
iD +|\sqrt{2}\sigma+\widetilde m_{j}|^2\,
\log{\frac{|\sqrt{2}\sigma+\widetilde m_{j}|^2}{\Lambda^2}}\right]\,.
\label{Veff}
\end{eqnarray}
This can be viewed as a master formula.

Equation (\ref{Veff}) presents exactly the effective potential which one would obtain from 
the HT model
(\ref{wcp}) by integrating out the $n^{wi}$ and $z^{wj}$ fields at large $N$ and $\tN$. As was expected,
the large-$N$
solutions of both models coincide.

\subsection{Switching on vacuum expectation values of $n$ and/or $z$}

Much in the same way as in the HT model, the strong coupling phase with the vanishing
vacuum expectation values  (VEVs) of both
$n$ and $z$ fields occurs in the $zn$ model at $m_i\sim m_j\sim\Lambda$ (we will discuss the vacuum structure
of the theory in the large-$N$ approximation in Sec. \ref{tvs}). 
At large/small masses the
fields $n$/$z$ develop VEVs and the theory is in the $n$-Higgs/$z$-Higgs phase, respectively.

To take into account the possibility of the $n$ and $z$ fields developing VEVs in (\ref{znferm}) we integrate out
 all $n$ and $z$ fields but one, say,  $n^1$ and  $z^1$, cf. \cite{Gorsky:2005ac}. 
 At the first stage this boils down to adding to (\ref{Sz}) the following term:
\beq
\int d^2 x \left(\left|\sqrt{2}\sigma+m_1\right|^2 +iD\right)\,|n^1|^2.
\label{n1term}
\eeq
At the second stage (integrating out $z'$s) we keep intact the terms depending on $z'^1$ in (\ref{Sz}).
This procedure leads us to the following final effective potential, which now depends
on the fields $\sigma$, $D$ and $n^1$, $z'^1$
\begin{eqnarray}
&&
V_{\rm eff}(\sigma,D,n^1, z'^1 )
\nonumber\\[3mm]
&=&\frac1{4\pi}
\sum_{i=2}^{N}\left[-\left(\Big|\sqrt{2}\sigma+m_i\Big|^2+iD\right)\,
\log{\frac{|\sqrt{2}\sigma+m_i|^2+iD}{\Lambda^2}} +iD 
\right.
\nonumber\\[3mm]
 &+&
\left.
|\sqrt{2}\sigma+m_i|^2\,
\log{\frac{|\sqrt{2}\sigma+m_i|^2}{\Lambda^2}}\right]
\nonumber\\[3mm]
&+&
\frac{1}{4\pi}\sum_{j=2}^{\tN}\left[-\left(\left|\sqrt{2}\sigma+
\widetilde m_{j}\right|^2-iD\right)\,
\log{\frac{|\sqrt{2}\sigma+\widetilde m_{j}|^2-iD}{\Lambda^2}}\right.
\nonumber\\[3mm]
 &-&
\left.
iD +|\sqrt{2}\sigma+\widetilde m_{j}|^2\,
\log{\frac{|\sqrt{2}\sigma+\widetilde m_{j}|^2}{\Lambda^2}}\right]
\nonumber\\[3mm]
 &+&
\left(\left|\sqrt{2} \sigma+m_1\right|^2 +iD\right)\,|n^1|^2 +\left(\left|\sqrt{2}\sigma+\widetilde{m}_1\right|^2 -iD\right)\,|z'^1|^2.
\label{Veffzn}
\end{eqnarray}
Varying the above expression with respect to the fields $\sigma$, $D$, $n^1$ and  $z'^1$ we derive the vacuum equations of the theory at large $N$, $\tN$.

\subsection{The vacuum structure}
\label{tvs}

Here we will briefly review the vacuum structure of the the HT and $zn$ models (for a detailed analysis see \cite{Koroteev:2010ct}). Given 
the fact that Eq.~(\ref{Veffzn}) is the same in both models, so are the solutions. 

First we shall consider the case of vanishing expectation values $\langle n^1\rangle$ and $\langle z'^1\rangle$ in Eq.~(\ref{Veffzn}), corresponding to the Coulomb branch of the theory. Then, due to relation \eqref{eq:VvsW}, the minima of the effective potential (\ref{Veffzn}) can be more easily extracted by determining the critical points of $\widetilde{W}_{{\rm eff}}$ (\ref{eq:ExactTwistSP}). In this way we then derive the following vacuum equation:\footnote{Note that this equation is valid for any $N$, not necessarily in the 't Hooft limit.}
\beq
\prod_{i=1}^{N}(\sqrt2\sigma + m_i)=\Lambda^{N-\widetilde N}\prod_{j=1}^{\widetilde N}(\sqrt2\sigma + \widetilde{m}_j)\,.
\label{vacuequ}
\eeq
Now, as was explained in Section \ref{Sec:Intro},
we choose the twisted masses in such a way that  the $\mathbb{Z}_N$ and $\mathbb{Z}_{\tilde N}$ discrete symmetries
are preserved, namely,\footnote{It is worth noting that a a generic choice of the twisted masses would  completely break  supersymmetry 
at the quantum level.} 
\beqn\label{eq:CircleMasses}
m_k  \eq   m \,e^{2\pi i\frac{k}{N}}\,,\quad k = 0,\dots, N-1\,,\nonumber\\[3mm]
\widetilde{m}_l  \eq   \widetilde{m}\, e^{2\pi i\frac{l}{\widetilde N}}\,,\quad l = 0,\dots, \widetilde N-1\,.
\eeqn
Then, Eq.~(\ref{vacuequ}) takes the following form:
\beq
(\sqrt2\sigma)^N +m^N= \Lambda^{N-\widetilde N}\left[(\sqrt2\sigma)^{\widetilde N} +\widetilde{m}^{\widetilde N}\right]\,.
\label{eq:exactvacua}
\eeq
The above equation obviously  has $N$ complex roots (assuming that $\tilde N < N$)
which can be easily found numerically for any $N$ and $\Nt$. Interestingly for large $N$ the solutions can be classified. 
For the future convenience we introduce a new parameter
\beq
\alpha = \frac{\tilde N}{N}\,,\qquad 0 < \alpha < 1\,.
\label{alp}
\eeq
Then, 
depending on the relation  between $\alpha, \, m$, and $\widetilde{m}$, there are two Coulomb branches, which are referred to as
\textbf{C}$m$ and \textbf{C}$\widetilde{m}$. The
roots of Eq.~(\ref{eq:exactvacua})  can be assigned to 
one of the following three groups:

\vspace{3mm}

\underline{$m$-{\em vacua}}:
In the domain \textbf{C}$m$, i.e.  $$\widetilde{m}<\Lambda \left(\frac{m}{\Lambda}\right)^{1/\alpha}\,,\quad m<\Lambda,$$  
\beq
\sqrt2\sigma_{m,l} =
  \Lambda \left(\frac{m}{\Lambda}\right)^{1/\alpha}e^{2\pi i \frac{l}{\widetilde N}}\,,   \quad \quad l=1,\dots,\Nt-1\,;
\label{eq:CoulombSmallVacua}
\eeq
\vspace{3mm}

\underline{$\Lambda$-{\em vacua}}:
These vacua exist only in the \textbf{C}$m$ domain and are located on the circle of radius $\Lambda$
\beq
\sqrt2\sigma_{\Lambda,k} = \Lambda\, e^{2\pi i \frac{k}{N-\widetilde N}}\,,\quad k = 0,\dots,N-\widetilde N-1\,;
\label{eq:CoulombLambda}
\eeq
\vspace{3mm}

\underline{$\widetilde{m}$-{\em vacua}}:
In the domain \textbf{C}$\widetilde{m}$, i.e. $$\widetilde{m}>\Lambda \left(\frac{m}{\Lambda}\right)^{1/\alpha}\,,\quad\widetilde{m}>\Lambda$$
\beq
\label{eq:Coulombm}
\sqrt2\sigma_{\widetilde{m},j}=\Lambda \left(\frac{\widetilde{m}}{\Lambda}\right)^{\alpha}e^{2\pi i \frac{j}{ N}} \quad \quad j=0,\dots,N-1\,.
\eeq
\vspace{3mm}

\begin{figure}[!ht]
\begin{center}
\includegraphics[height=10cm, width=10cm]{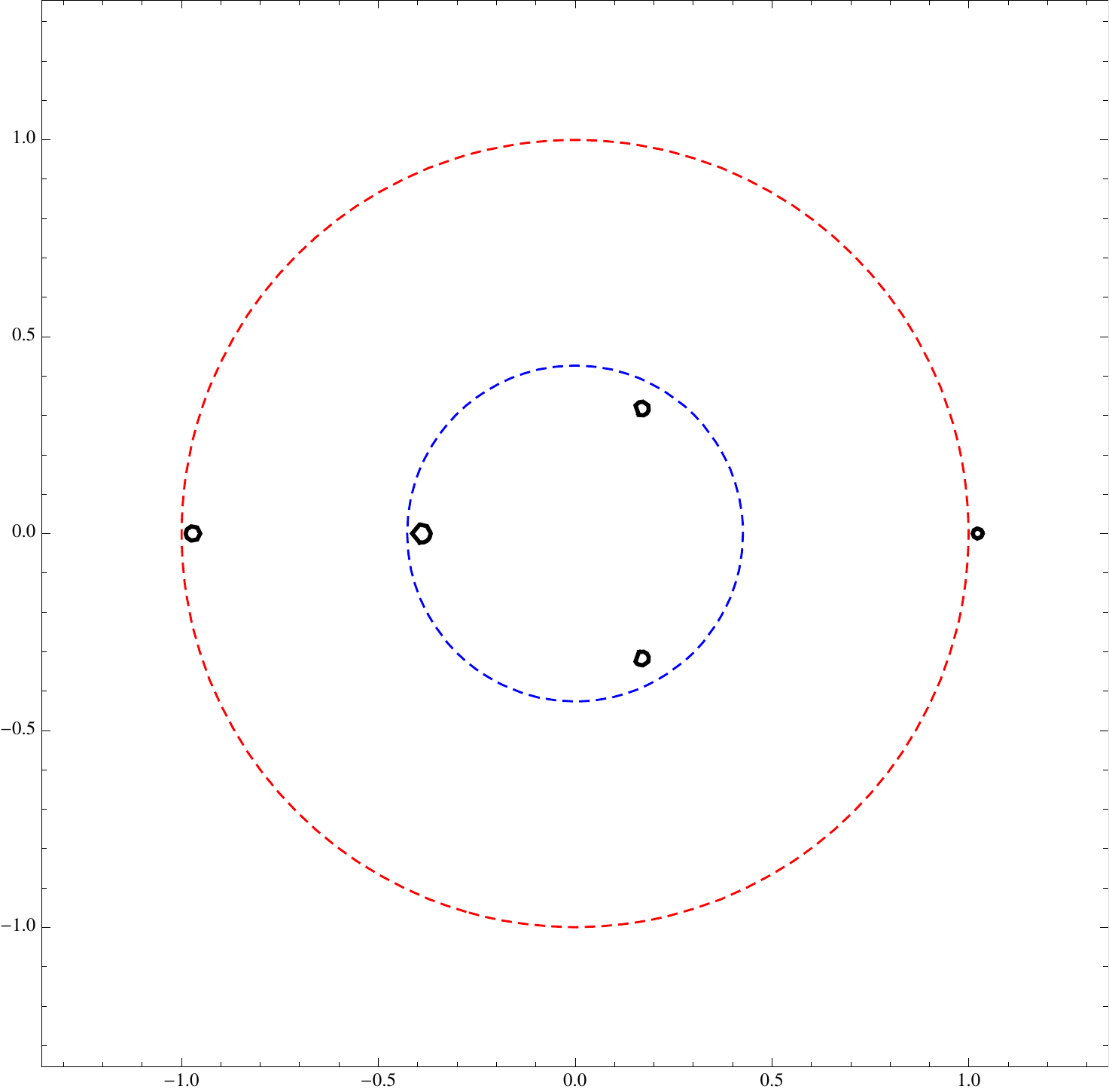}
\caption{Vacua of the HT model for $N=5, \Nt=3$ in the \textbf{C}$m$ domain. We can see two ($N-\Nt=2$) $\Lambda$-vacua near the circle of radius $\Lambda$ and three ($\Nt=3$) $m$-vacua near the circle or radius $m^{1/\alpha}$ in units of $\Lambda$.}
\label{fig:vacua5v3}
\end{center}
\end{figure}

\begin{figure}[!h]
\begin{center}
\includegraphics[height=10cm, width=10cm]{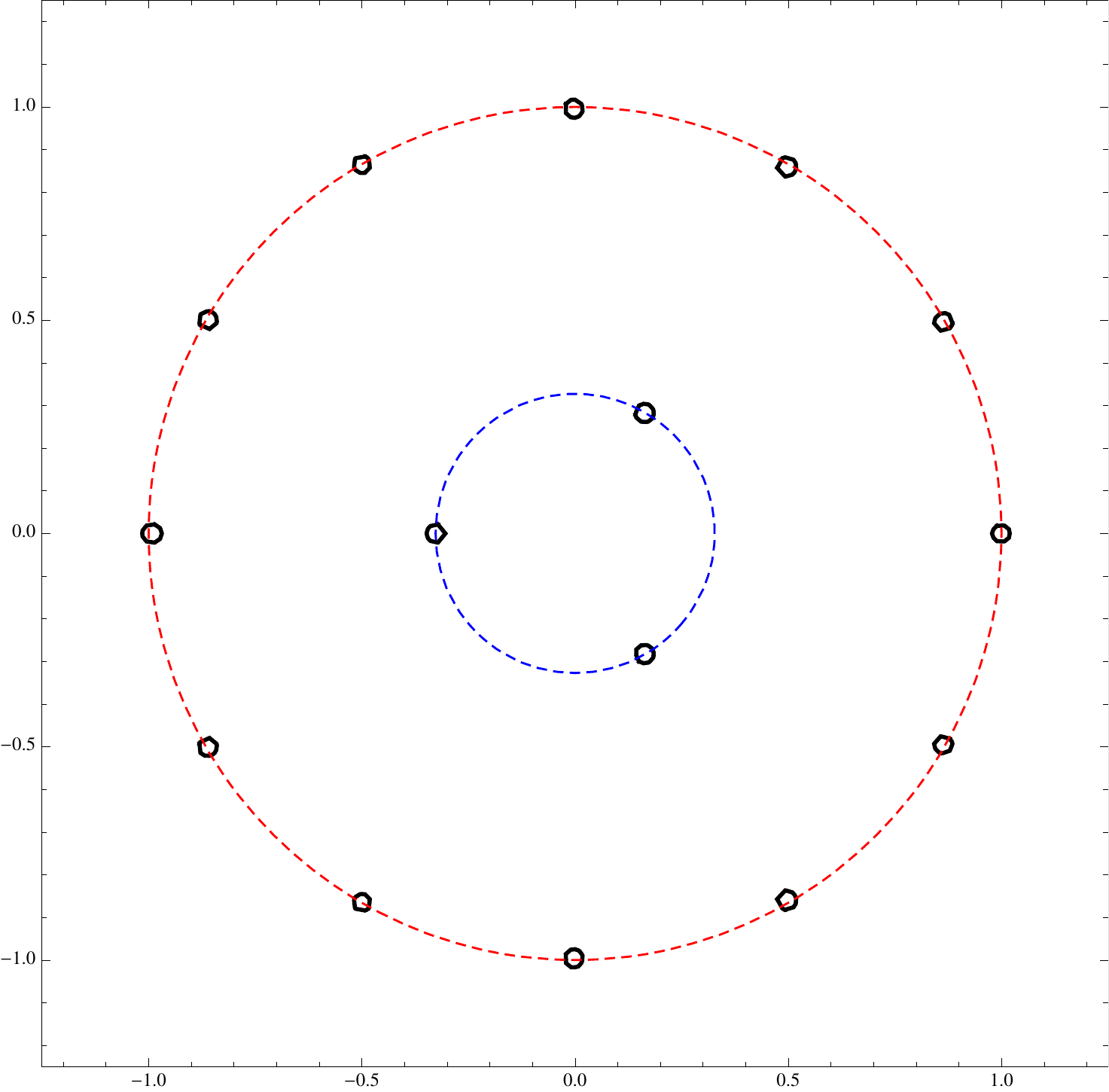}
\caption{Vacua of the HT model for $N=15, \Nt=3$ in the \textbf{C}$m$ domain. For larger values of $N$ the formulae (\ref{eq:CoulombSmallVacua}) and (\ref{eq:CoulombLambda}) are getting more precise. Small circle has radius $m^{1/\alpha}$ in units of $\Lambda$.}
\label{fig:vacua15v3}
\end{center}
\end{figure}

\begin{figure}[!h]
\begin{center}
\includegraphics[height=10cm, width=10cm]{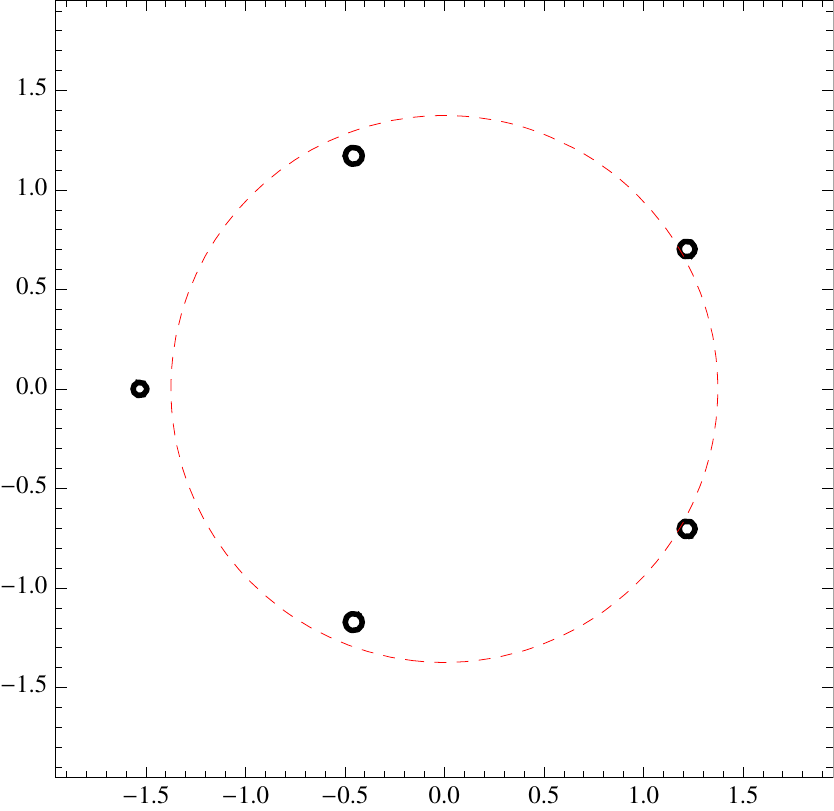}
\caption{Vacua of the HT model for $N=5, \Nt=3$ in the \textbf{C}$\widetilde{m}$ domain. All vacua localize near the circle  of radius $\widetilde{m}^\alpha$ in units of $\Lambda$.}
\label{fig:vacua5}
\end{center}
\end{figure}

The above expressions are approximate to the leading order in $1/N$. For small $N$ there will be corrections, see Figs.~\ref{fig:vacua5v3}, \ref{fig:vacua15v3}, and \ref{fig:vacua5}.  These figures depict
the  complex $\sigma$ plane; the actual vacua that solve Eq.~(\ref{eq:exactvacua}) are located at the centers of 
the small black nodes in these figures,  while the dashed circles drawn for reference have radii given by Eqs.~(\ref{eq:CoulombSmallVacua}),
(\ref{eq:CoulombLambda}), and (\ref{eq:Coulombm}). 

Note also that in the  regime
\beq
\frac{\widetilde{m}}{\Lambda}=\left(\frac{m}{\Lambda}\right)^{1/\alpha}\,,
\label{eq:LineSCFT}
\eeq
 Eq.~(\ref{eq:exactvacua}) degenerates into 
\beq
(\sqrt2\sigma)^N = \Lambda^{N-\widetilde N}(\sqrt2\sigma)^{\widetilde N}\,.
\label{eq:exactvacuared}
\eeq
This equation has two sets of solutions,
\beq
(\sqrt2\sigma)^{N-\widetilde N} = \Lambda^{N-\widetilde N}, \quad \sigma=0\,,
\eeq
where the former solution gives $N-\Nt$ massive vacua and the latter applies to the conformal regime.

There are two Higgs branches corresponding to $\langle n^1\rangle\neq 0$ and $\langle z'^1\rangle \neq 0$ in \eqref{Veffzn}. The former exists for $m>\Lambda$ and $m/\Lambda>(\widetilde{m}/\Lambda)^\alpha$ whereas the conditions for the latter are $(m/\Lambda)^{1/\alpha}<\widetilde{m}/\Lambda<1$. If $n^1$ or $z'^1$ develop VEVs we must work with Eq.~(\ref{Veffzn}), minimizing $V_{{\rm eff}}$. This minimization was done in \cite{Koroteev:2010ct} and we refer the reader to this paper for further details.

\subsection{Non-BPS spectrum}
\label{Sec:Spectrumzn}

\begin{figure}[!h]
\begin{center}
\includegraphics[height=2.4cm, width=10.5cm]{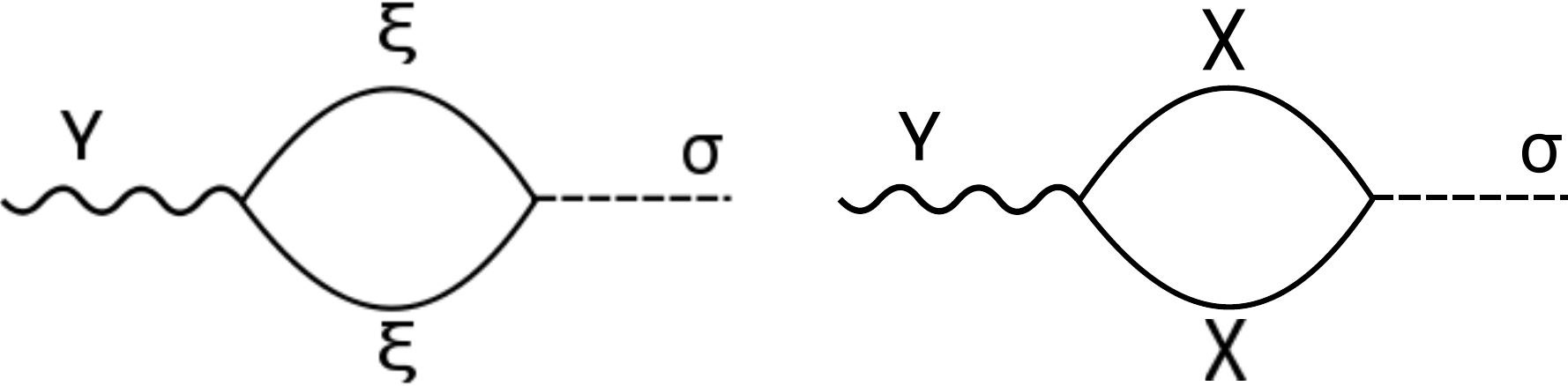}
\caption{One-loop diagrams which contribute to the the photon-scalar anomalous mixing.}
\label{fig:PhotonScalarMixingWeighted}
\end{center}
\end{figure}

In Sec.~\ref{Sec:twisted} 
we demonstrated
that the spectrum of the $zn$ model in the large-$N$ limit coincides with that of the HT model;
the latter was discussed in detail in \cite{Koroteev:2010ct}. Here we will calculate the mass  of the particles from
the vector multiplet $V$. As was discussed
 above, there are $N-\Nt$  $\Lambda$-vacua in this model. Let us choose  form Eq.~(\ref{eq:CoulombLambda})
 the real vacuum, namely, 
 \beq
 \sqrt2\sigma_0=\Lambda
 \eeq
 and consider field fluctuations around this vacuum (all $\Lambda$-vacua are physically equivalent). The effective action 
 for these fluctuations is 
\beqn
\label{eq:Lagr1loopeff}
\cell &=&
 -\frac{1}{4e_\gamma^2}F_{\mu\nu}^2 + \frac{1}{e_{\sigma\,1}^2}(\dpod{\mu}\mathfrak{Re}\,\sigma)^2 + \frac{1}{e_{\sigma\,2}^2}(\dpod{\mu}\mathfrak{Im}\,\sigma)^2 + i\frac{1}{e_\lambda^2}\,\bar{\lambda}\gamma^\mu\nabla_\mu \lambda 
\nonumber\\[3mm]
&+&
i\mathfrak{Im}(\bar{b}\,\sigma)\, \epsilon_{\mu\nu}F^{\mu\nu}   -V_{\text{\rm eff}}(\sigma)-(i\Gamma \bar{\sigma}\bar{\lambda}\lambda+\text{H.c.})\,.
\eeqn
In the above formula the effective potential $V_{\rm eff}(\sigma)$ is given by Eq.~(\ref{Veff}),
while
the gauge and scalar couplings can be calculated from the corresponding one-loop Feynman diagrams. 
The gauge field is coupled to the imaginary 
part of $\sigma$. Figure~\ref{fig:PhotonScalarMixingWeighted} displays the one-loop diagrams which contribute to the mixing. All relevant calculations were carried out in \cite{Koroteev:2010ct}. Here, in addition to these results, we find the mass of the photon  from the vector multiplet.
%

\paragraph{Masses.}

For vanishing twisted masses the one-loop superpotential Eq.~(\ref{Veff}) takes the following form:
\vspace{2mm}
\beqn
\label{eq:V1loop}
V_{1-{\rm loop}} 
&=&
 \frac{N}{4\pi}\left(-\left(iD+2\left|\sigma\right|^2\right)\log\frac{2\left|\sigma\right|^2+iD}{\Lambda^2}+2\left|\sigma\right|^2\log\frac{2\left|\sigma\right|^2}{\Lambda^2}\right)
\nonumber\\[3mm]
&-&
\frac{\Nt}{4\pi}\left(-\left(iD-2\left|\sigma\right|^2\right)\log\frac{2\left|\sigma\right|^2-iD}{\Lambda^2}-2\left|\sigma\right|^2\log\frac{2\left|\sigma\right|^2}{\Lambda^2}\right)
\nonumber\\[3mm]
&+&
\frac{N-\widetilde N}{4\pi}iD\,.
\eeqn
In the case of vanishing twisted masses we can approximately solve the vacuum equation on the Coulomb branch,
\beq
N\log\frac{2|\sigma|^2+iD}{\Lambda^2}-\Nt\log\frac{2|\sigma|^2-iD}{\Lambda^2}=0\,.
\eeq
Near the vacuum $\sqrt{2}\sigma =\Lambda$ we expect $D$ to be small.
Therefore, we can rewrite the above equation as
\beq
N\log\left(1+\frac{iD}{2|\sigma|^2}\right)-\Nt\log\left(1-\frac{iD}{2|\sigma|^2}\right)+
(N-\Nt)\log\frac{2|\sigma|^2}{\Lambda^2}=0\,.
\eeq
Then, Taylor-expanding and denoting 
\beq
d=\frac{iD}{\Lambda^2}\,, \qquad s= \frac{2\mathfrak{Re}(\sqrt{2}\sigma-\Lambda)}{\Lambda}\,,
\eeq
we get
\beq
\label{eq:dvss}
d=-\frac{N-\Nt}{N+\Nt}\,s\,.
\eeq
Equation~(\ref{eq:V1loop}) can be rewritten  in terms of new variables as  
\beqn
V_{1-loop} 
&=&
 \frac{N\Lambda^2}{4\pi}\Bigg[-s (\alpha -1)-(s+1) (\alpha +1) \log (s+1)
\nonumber\\[3mm]
&+&
\frac{(2 s \alpha +\alpha +1)}{\alpha +1} \log \left(\frac{2 s
   \alpha +\alpha +1}{\alpha +1}\right)
  \nonumber\\[3mm] 
  &+&
 \frac{\alpha  (2 s+\alpha +1) }{\alpha +1}\log
   \left(\frac{2 s}{\alpha +1}+1\right)\Bigg]\,,
\eeqn
where  $\alpha $ is defined in Eq.~(\ref{alp}). Using Eq.~(\ref{eq:dvss}) we get, to the second order in $s$,
\beq
V_{1-{\rm loop}}=\frac{ (N-\Nt)^2}{2(N+\Nt)}\frac{s^2 \Lambda ^2}{4\pi}\,.
\eeq

Next, we will  canonically normalize the kinetic terms in Eq.~(\ref{eq:Lagr1loopeff}). In particular,
 we do a rescaling $$\mathfrak{Re}\,\sigma \to e_{\sigma\, 1}(\mathfrak{Re}\,\sigma)\,.$$
 As was shown in \cite{Koroteev:2010ct}
\beq
e_{\sigma\,1}^2=\frac{4\pi}{(N-\Nt)\Lambda^2}\,.
\eeq
Therefore, the mass of real part of sigma is 
\beq
m_{\sigma_1} = 2\sqrt{\frac{N-\Nt}{N+\Nt}}\Lambda\,.
\eeq
Note that this expression has $1/N$ corrections since the vacua (\ref{eq:CoulombLambda}) are given to the leading order in $N$.
Due to supersymmetry the masses of the photon, fermion and scalar fields are equal,
\beq
m_\gamma = m_\lambda = m_\sigma\,.
\eeq

Notice that, as should be obvious from the discussion in Section \ref{Sec:beta}, and as we confirmed in this section with an explicit calculation, the full spectra of the $zn$ model and the HT model, including the non-BPS sector, are equivalent  at the leading order in the large-$N$ approximation.

\section{$\NLSM$ Description and Geometric Renormalization}\label{Sec:GeometricFormulations}

As was first observed in Ref.  \cite{Shifman:2011xc}, the HT and $zn$ models 
have different metrics on their respective vacua manifolds. In this section we will
investigate perturbation theory of both models using a nonlinear sigma model ($\NLSM$) description. We will consider in parallel the geometry of the  $zn$ and HT models and study their one-loop renormalization in the geometric language. We will also  show that the K\"{a}hler potential of the HT model reduces to that of the $zn$ model in a {\em certain limit}.

\paragraph{From $\GLSM$ to $\NLSM$.}

Let us first illustrate the main idea with a simple example.  We will review here how a vacuum manifold of the $\CP{1}$ $\NLSM$   emerges from the gauged description of the model in the limit when the gauge coupling(s) are sent to infinity.  

The corresponding gauged linear sigma model
($\GLSM$) Lagrangian for the $\CP{1}$ model  in the superfields formalism reads
\[\label{eq:LagrCP1}
\lagr =\int d^4\theta\,\left( \left(|X_1|^2+|X_2|^2\right)e^V - r V +\frac{1}{e^2}|\Sigma|^2\right)\,,
\]
where $X_1, X_2$ are chiral multiplets, $V$ is a  twisted vector multiplet with field strength $\Sigma$, $r$ is the FI parameter, and $e$ is the gauge coupling. One can see that the following term belongs to the Lagrangian:
\[\label{eq:DtermCP1}
D(|x_1|^2+|x_2|^2-r)\,,
\]
which gives rise to the D-term constraint and it  comes from the terms linear in $V$. Here $x_{1,2}$ are the bottom components of fields $X_{1,2}$. The constraint modulo the U$(1)$ symmetry $(\Complex^2-Z)//U(1)$, where $Z$ is the locus of $|x_1|^2+|x_2|^2-r$ defines the vacuum target manifold of the model. In this particular case is given by $\CP{1}\simeq S^2$, the two-dimensional sphere of radius $r$. By making the radius of the sphere very large we go into the flat limit and the target manifold of the model should simply reduce to  $\Complex^1$. However, this statement is not evident from analyzing the D-term constraint \eqref{eq:DtermCP1}. The reason for this is that $X_1$ and $X_2$ are not the true coordinates of the vacuum manifold, but their ratio is. Indeed, integrating out $V$ in \eqref{eq:LagrCP1} we get
\[
\lagr = \int d^4\theta\, r\log \left(|X_1|^2+|X_2|^2\right)\,.
\]
Now we need to fix the gauge in order to keep only physical degrees of freedom, doing this we obtain the K\"{a}hler potential for the $\CP{1}$ model
\[\label{eq:KahPotCP1}
K = r\log (1+|X|^2)\,,
\]
where $X=X_2/X_1$. Let us further do the rescaling $X\to X/\sqrt{r}$ and take the limit $r\to+\infty$. What we get is 
\[
K = |X|^2\,,
\]
which corresponds to the flat metric on $\Complex$. Note that one could have considered \eqref{eq:KahPotCP1} and instead of doing the rescaling expand the K\"{a}hler potential for fixed $r$ at small values of $|X|^2$ and get the same result. It is, of course, a reflection of the equivalence of rescaling the coordinates and metric. We will compare the HT and $zn$ models later in this section using small field expansion. In the following subsections we will get the vacuum manifolds for the two models in question from their gauged descriptions which have been reviewed in Sec.~\ref{Sec:HananyTongModel}.

\subsection{The $zn$ model vs. the HT model}\label{theznm}

The following Lagrangian describes the  $zn$ model \cite{Shifman:2011xc}
\[\label{eq:LagrZNeInf}
\lagr_{zn} = \int d^4\theta\, \left(|\mathcal{N}_i|^2 \me^V + |\mathcal{Z}_j|^2|\mathcal{N}_i|^2-r V +\frac{1}{e^2}|\Sigma|^2\right)\,,
\]
where we use the following chiral superfields
\<
\mathcal{N}^i \eq n^i + \sqrt{2}\theta^\alpha \xi^i_\alpha + \bar{\theta}\theta F^i\,,\quad i=1,\dots, N \nln
\mathcal{Z}^j \eq z^j + \sqrt{2}\theta^\alpha \chi^j_\alpha  + \bar{\theta}\theta \widetilde{F}^j\,,\quad j=1,\dots, \Nt\,,
\>
vector field $V$ in the Wess--Zumino gauge ($\theta^1 = \theta^+\,,\theta^2=\theta^-$ and the same for dotted components, see \cite{Witten:1993yc})
\begin{align}
V &= \theta^+\bar{\theta}^+(A_0+A_3)+\theta^-\bar{\theta}^-(A_0-A_3)+
i\sqrt{2}\sigma\theta^-\bar{\theta}^+ 
+i\sqrt{2}\bar{\sigma}\theta^+\bar{\theta}^- \notag\\
&+\left(2i\theta^-\theta^+(\bar{\theta}^-\bar{\lambda}_- +\bar{\theta}^+\bar{\lambda}_+)+\text{H.c.}\right)+\half \theta^4 D\,,
\end{align}
and the twisted chiral field $\Sigma = \mathcal{D}_+\bar{\mathcal{D}}_- V$ 
\[
\Sigma = \sigma + i\sqrt{2}\theta^+\bar{\lambda}_+ -i\sqrt{2} \bar{\theta}^-\lambda_- + \theta^+\bar{\theta}^- (D-i F_{01})\,.
\]
Given the above superfield representations one can derive the full action of the $zn$ model in components \eqref{znferm}.

\paragraph{Vacuum manifold of the $zn$ model.}

Let us proceed with the geometric description of the theory. Taking the limit $e\to\infty$ and integrating out vector superfield $V$ in \eqref{eq:LagrZNeInf} we arrive at 
the following Lagrangian:
\<
\lagr_{zn} = \int d^4\theta\, \left(|\mathcal{Z}_j\,\mathcal{N}_i|^2+r \log |\mathcal{N}_i|^2\right)\,.\>
Similarly to the $\CP{1}$ case described above we need to get rid of the unphysical degree of freedom which is present in the above expression.
If we define\,\footnote{Assuming $\mathcal{N}_N\neq 0$.}
\beqn
\label{eq:ChangeVarCP}
\Phi_i 
&=&
 \frac{\mathcal{N}_i}{\mathcal{N}_N}\,, \quad i=1,\dots ,N-1\,,\nonumber\\[3mm]
\mathfrak{z_j} &=& r^{-1/2} \mathcal{N}_N \mathcal{Z}_j\,, \quad j=1,\dots ,\Nt\,,
\eeqn
we get the following K\"{a}hler potential for the $zn$ model:
\[\label{eq:znKahlerPotential}
K_{zn} =r|\zeta|^2 + r\log (1+|\Phi_i|^2)\,,
\]
where 
\[
\label{sh}
|\zeta|^2 \equiv |\mathfrak{z}_j|^2(1+|\Phi_i|^2)\,.
\]
Note that $\zeta$ is not a holomorphic variable in any sense. We  use the  notation (\ref{sh}) as a shorthand. $|\zeta|^2$ is the only combination involving
$\mathfrak{z_j} $'s which is invariant under the global symmetries (\ref{globalsym}) of the model. 
Needless to say, so is any power of $|\zeta|^2$.

The K\"{a}hler potential \eqref{eq:znKahlerPotential} describes geometry of the vacuum manifold of the $zn$ model in terms of $(N+\tN -1)$ unconstrained complex variables. The global SU$(N)$ is realized nonlinearly much in the same way as in the $\CP{N-1}$ model while the SU$(\Nt)$ symmetry is realized linearly on the $\mathfrak{z}_j$ fields.  For $\tN=1$, the K\"{a}hler potential (\ref{eq:znKahlerPotential}) reduces to that describing the {\em blow-up} of the $\mathbb{C}^N$ space at the origin \cite{MR962489}. In this case we can observe that the SU$(N)$ symmetry becomes manifest and is realized as the isometry of the target space after the following redefinition:
\[
|\zeta|^2 = |\Xi_i|^2\,,\quad \Xi_1 = \mathfrak{z}_1\,,\quad \Xi_i = \mathfrak{z}_1\Phi_i\,,\quad i=2,\dots,N\,.
\]
In this case the K\"{a}hler potential takes the form
\[
K_{zn} =r|\Xi_i|^2 + r\log |\Xi_i|^2\,.
\]

It is instructive to reiterate to make explicit all isometries of \eqref{eq:znKahlerPotential}. For simplicity we put $N=1$, so that the second part of the action in \eqref{eq:znKahlerPotential} is, in fact,  that of $\CP{1}$. As is well known, $\CP{1}$ is invariant under nonhomogenious nonlinear transformations
\beqn
\Phi \to \Phi + \beta + \bar\beta \, \Phi^2\,, \qquad  \bar{\Phi} \to \bar{\Phi} + \bar\beta + \beta \,\bar{\Phi}^2\,,
\label{su2u1}
\eeqn
where $\beta$ and $\bar\beta$ are infinetissimal transformation parameters. This expresses the ${\rm SU}(2)/{\rm U}(1)$ invariance of the $\CP{1}$ action.
Indeed, under these transformations
\beq
1+ \Phi\bar{\Phi} \to \left(1+ \Phi\bar{\Phi}\right)\left(1+ \beta\bar{\Phi}\right)\left(1+ \bar\beta \, \Phi\right)
\eeq
implying K\"ahler transformations of $\log\left(1+|\Phi|^{2}\right)$ under which the $\CP{1}$ action is invariant.  Let us supplement (\ref{su2u1}) by the following holomorphic transformations of the variables $\mathfrak{z}_j$
\beq
\mathfrak{z}_j \to \frac{\mathfrak{z}_j }{1+ \bar\beta \, \Phi}\,,\qquad \bar{\mathfrak{z}}_j \to \frac{\bar{\mathfrak{z}}_j }{1+ \beta \, \bar{\Phi}}\,.
\label{su2u1p}
\eeq
We immediately confirm that $ |\zeta|^{2} $ is invariant under the combined action of (\ref{su2u1}) and (\ref{su2u1p}). 
Here it is obvious that
this is the only independent invariant of this type. Thus the observed symmetry only allows polynomials in $|\zeta|^2$ in the K\"ahler potential.

\paragraph{Vacuum manifold of the HT model.}

Using the same notations for the superfields as for the $zn$ model we can formulate the HT model \eqref{wcp} as the following $\GLSM$ ($e\to\infty$):
\[\label{eq:LagrHTeInf}
\lagr_{\text{HT}} = \int d^4\theta\, \left(|\mathcal{N}_i|^2 \me^V + |\mathcal{Z}_j|^2\me^{-V}-r V\right)\,.
\]
Using the same change of variables as in \eqref{eq:ChangeVarCP}, after integrating out $V$ in \eqref{eq:LagrHTeInf} we obtain the 
K\"{a}hler potential for the HT model,
\[
K_{\text{HT}} = \sqrt{r^2+4r|\zeta|^2}- r \log \left(r+\sqrt{r^2+4r |\zeta|^2}\right)+ r \log(1+|\Phi_i|^2)\,.
\label{eq:HTpotentialred}
\]
For $N=2,\,\Nt=1$, the  K\"{a}hler potential (\ref{eq:HTpotentialred}) describes the so-called Eguchi--Hanson space and was discovered by Calabi \cite{MR543218}. For generic $\Nt$ the target manifold in question is the $\mathcal{O}(-1)^{\Nt}$ tautological fiber bundle over $\CP{N-1}$. For a
mathematical derivation of the K\"{a}hler potential \eqref{eq:HTpotentialred} see \cite{2010arXiv1004.4049L}.

\paragraph{From the HT model to the $zn$ model.}

At first sight the $zn$ and HT models look quite different, as much as  their K\"{a}hler potentials  (\ref{eq:znKahlerPotential}) and  (\ref{eq:HTpotentialred}). 
This is indeed the case, but there is a domain of the target space  where they reduce to the same model. As we have already mentioned, the target manifold of the HT model is the total space of the $\Nt$-th power of the tautological bundle over $\CP{N-1}$. Thus this is a noncompact manifold with $\Nt$ noncompact directions. 

We will now make a more quantitative comparison of the two models. Let us consider Eq.~(\ref{eq:HTpotentialred}) at small values of $|\zeta|^2$.  The result of the small $|\zeta|^2$-expansion depends on the sign of the FI parameter $r$. Below we will consider both branches.

\vspace{3mm}

\noindent
(i) $r>0$: \hspace{3mm}
For small $|\zeta|^2$ we can Taylor-expand around $|\zeta|^2=0$  and observe that the K\"{a}hler potential 
(\ref{eq:HTpotentialred}) in the second order in $|\zeta|^2$ takes the form
\beq
\label{eq:KahlerPotrl0dual}
K_{\text{HT}} =r|\zeta|^2 + r\log (1+ |\Phi_i|^2)+\mathcal{O}(|\zeta|^4)\,,
\eeq
This K\"{a}hler potential coincides with  the one (\ref{eq:znKahlerPotential}) of the $zn$ model.

\vspace{3mm}

\noindent
(ii) $r<0$: \hspace{3mm}
Small-$|\zeta|^2$ expansion gives the following K\"{a}hler potential:
\beq\label{eq:KahlerPotrb0}
K_{\text{HT}} =r|\zeta|^2 - r\log (1+ |\widetilde{\mathfrak{z}}_j|^2)+\mathcal{O}(|\zeta|^4)\,,
\eeq
where\,\footnote{Again, it assumed that $\mathcal{Z}_{\Nt}\neq 0$.}
\[
\widetilde{\mathfrak{z}}_j = \frac{\mathcal{Z}_j}{\mathcal{Z}_{\Nt}}\,\quad j=1,\dots, \Nt-1.
\]
This model corresponds to the dual $zn$ sigma model with $\CP{\Nt-1}$ as the base manifold. One can rewrite its K\"{a}hler potential as follows
\[
K_{\widetilde{zn}} =r |\mathcal{N}_i|^2(1+|\widetilde{\mathfrak{z}}_j|^2)+\widetilde{r}\log (1+ |\widetilde{\mathfrak{z}}_j|^2)\,,
\]
where $\widetilde{r}=-r>0$. This manifold has $N$ noncompact and $\Nt$ compact directions. As we will see later,
the one-loop $\beta$ function (or first Chern class of the bundle) will be proportional in this case to $N-\Nt$. Once we start with the HT model \eqref{eq:HTpotentialred} with $N>\Nt$, corresponding to the asymptotically free theory, $N-\Nt$ will be positive, which will entail growth of the FI parameter $\widetilde{r}$ along the RG flow. Thus the dual $zn$ model is not asymptotically free, but rather IR free. In what follows we will only concentrate on the first case $r>0$.

Thus far we considered small values of $|\zeta|^2$. On the contrary, at large values of $|\zeta|^2$, as can be seen from Eqs.~(\ref{eq:znKahlerPotential}) and  (\ref{eq:HTpotentialred}), the two models behave differently. As was shown in \cite{Shifman:2011xc}, 
 in this limit the $zn$ model has vanishing scalar curvature, whereas the HT model has not.

One can see from \eqref{eq:KahlerPotrl0dual} that in the leading order the HT and $zn$ models have the same K\"{a}hler potential,
\[
K_{\text{HT}} = K_{zn} + \mathcal{O}(|\zeta|^2)\,.
\]
  This observation suggests that at one loop, in the leading order in $|\zeta|^2$ the two models have the same one-loop $\beta$ functions. Nevertheless, beyond 
one loop one expects the theories to have different $\beta$ functions. Moreover, even at one loop for large values of $|\zeta|^2$ the two models get different corrections. We will give explicit expressions later on in this section.

\subsection{Perturbation theory}

For any K\"{a}hler nonlinear sigma model with the K\"{a}hler metric $g_{i\bar{\jmath}}$ and coupling constant $g$ the 
Gel-Mann--Low functional (in what follows we shall call it $\beta$ function for short) reads \cite{Fradkin:1985ys,Metsaev:1987bc,Foakes:1987gg,Graham:1987ep,Jack:1988sw}
\beq\label{eq:PertSeries}
\beta_{i\bar{\jmath}} =  a^{(1)} R^{(1)}_{i\bar{\jmath}} + \frac{1}{2r}a^{(2)}R^{(2)}_{i\bar{\jmath}} + \dots\,,
\eeq
where $a^{(k)}$ are some constants  ($k=1,2, ...)$ and $R^{(k)}$ are operators composed from $k$-th power 
of the curvature tensors (see e.g. (\ref{eq:TableBetaCoeffs})). According to the above series a contribution from the $n$th loop scales as $r^{1-n}$. For the metric of a general  form the first several terms are known. The first two of them are
\beqn
\label{eq:TableBetaCoeffs}
R^{(1)}_{i\bar{\jmath}}\eq R_{i\bar{\jmath}}\,,\nonumber \\[2mm]
R^{(2)}_{i\bar{\jmath}}\eq R_{i\bar{k}l\bar{m}}R^{\bar{k}\,\,\,l\bar{m}}_{\,\,\bar{\jmath}}\,.
\eeqn
In supersymmetric sigma models, however, most of the coefficients $a^{(k)}$ from \eqref{eq:PertSeries} vanish. For example, in supersymmetric $\CP{N-1}$ sigma model all terms except the first one in \eqref{eq:PertSeries} are zero \cite{Morozov:1984ad}. The calculation was based on the instanton counting \cite{Novikov:1983ee} and the coefficients of the $\beta$ function were expressed in terms of the number of the zero modes.
 
The common lore in perturbation theory of nonlinear sigma models suggests that for generic K\"ahler manifolds the theory is nonrenormalizable, as each order in the perturbation series (\ref{eq:PertSeries}) brings in a new operator, with a different field dependence. For some particular symmetric target manifolds e.g. for the Einstein manifolds, no new structures are produced.  The renormalization is merely reduced to a single coupling constant renormalization. It is easy to see that the HT and $zn$ model target spaces are not of this kind and all terms in the series 
(\ref{eq:PertSeries}) have different field dependence. However, let us have a closer look the one-loop perturbation theory and see how we can deal with the above mentioned nonrenormalizability.

\paragraph{One-loop renormalization of the K\"{a}hler potential in the $zn$ model.}

\vspace{3mm}

For a K\"{a}hler manifold with the K\"{a}hler potential $K(z_i,\bar{z}_i)$ the metric is given by
\beq
g_{i\bar{\jmath}} = \partial_i\bar{\partial}_{\bar{\jmath}}K(z_i,\bar{z}_i)\,,
\eeq
while all  other components (such as $g_{ij}=0$) vanish. The corresponding Ricci tensor is therefore a total derivative and is given by 
\beq\label{eq:RicciKahler}
R_{i\bar{\jmath}}=-\partial_i\bar{\partial}_{\bar{\jmath}}\log \text{det}(g_{i\bar{\jmath}})\,.
\eeq
For Einstein manifolds Ricci tensor is proportional to the metric, therefore 
\[
-\log\text{det}(g_{i\bar{\jmath}})= \alpha K(z_i,\bar{z}_i)\,
\]
up to a K\"{a}hler transformation. For instance, for the $\CP{N-1}$ model the coefficient $\alpha$ in the above formula is equal to $N$. As we emphasized previously, for the $\CP{N-1}$ model this result is exact:  higher loops do not give any corrections to the $\beta$ function. 

Let us now examine the curvature tensors for the $zn$ model. It turns out that the calculation of the determinant of the metric tensor can be performed exactly for any $N$ and $\Nt$; the answer is more intricate in the HT model. After some calculations we get\footnote{This result holds up to an additive constant which depends on $r$. Since the Ricci tensor is a total derivative we can allow such a freedom. Certainly we can also change this expression by a K\"{a}hler transformation.}
\[\label{eq:logdetznfull}
-\log\text{det}(g^{(zn)}_{i\bar{\jmath}})= (N-\widetilde{N})\log(1+|\Phi_i|^2)-(N-1)\log(1+|\zeta|^{2})\,.
\]
Let us at this point derive the same quantity for the HT model in order to show how its one-loop result deviates from the one for the $zn$ model. For the HT model a generic formula is harder to get, we therefore focus on an example for, say, $N=2,\,\Nt=1$. One gets
\[\label{eq:HTRicci1loop21}
-\log\text{det}(g^{(\text{HT})}_{i\bar{\jmath}})= \log(1+|\Phi_i|^2) - \log\left(1+\frac{r}{\sqrt{r^2+4r|\zeta|^2}}\right)\,.
\]
This expression obviously gives a different correction to the K\"{a}hler potential.

Formula  \eqref{eq:logdetznfull} means that the K\"{a}hler potential acquires an infinite correction and becomes
\[\label{eq:KahlerPotzn1loop}
K^{(1)}_{zn} = \left(r_0-\frac{N-\Nt}{2\pi}\log\frac{M}{\mu}\right)\log(1+|\Phi_i|^2)+|\zeta|^2+\frac{N-1}{2\pi}\log\frac{M}{\mu}\log(1+|\zeta|^{2})\,,
\]
where $M$ is the UV cutoff and $\mu$ is the normalization scale. We immediately see that the target manifold of the $zn$ model is not of the 
Einstein type. We can also see that in order to eliminate the divergence in the last term in the above formula one has to introduce a new counterterm.

\paragraph{A side remark.}

There exist the so-called {\em quasi-Enstein} manifolds or {\em Ricci solitons}, for which the following equality takes place:
\[
R_{i\bar{\jmath}}=\alpha g_{i\bar{\jmath}}+\partial_i \bar{v}_{\bar{\jmath}}+\bar{\partial}_{\bar{\jmath}} v_i\,
\]
for some vector field $v$. None of the manifolds considered in this paper are of this kind. Indeed, one can check that K\"{a}hlerian structure imposes constraints on the vector field $v$ which are not satisfied in either $zn$ or HT models. Quasi-Einstein K\"{a}hler manifolds had been investigated by a number of mathematicians as well as physicists. It was shown by Friedan \cite{Friedan:1980jf, Friedan:1980jm}, who carried out  a stability analysis of RG equations at one loop, that a fix point of the RG flow has to be a quasi-Einstein manifold. Quasi-Einstein manifolds are quite hard to find explicitly, for most of the known cases their K\"{a}hler potentials are known only implicitly and in quadratures (see e.g. \cite{2010arXiv1004.4049L} and references therein for examples related to our work). However, in the special case of $N=\Nt=1$ one can specify the metric explicitly. Its only nonzero component is given by (see \cite{2010arXiv1004.4049L})
\[\label{eq:RSFriedan11}
g_{RS} = \frac{r}{1+|z|^2}\,,
\]
where $z$ is a coordinate on the target manifold. Note that for $N=\Nt=1$ the $zn$ model is trivial: it has $\Complex$ as its target space, whereas the HT model has
a nontrivial
 metric\,\footnote{Note that this metric appears on the Higgs branch of the theory when two twisted masses collide (the Argyres--Douglas point) \cite{Hanany:1997vm}. The space is asymptotically $\Complex/\Integers_2$.}
\[
g_{\text{HT}} = \frac{r}{\sqrt{1+|z|^2}}\,.
\]
Based on the arguments given in \cite{Friedan:1980jf, Friedan:1980jm} the HT model in this case should flow to the space with metric \eqref{eq:RSFriedan11}. Studying the fixed points of the RG flow in $\NLSM$s is an interesting question, but it is beyond the scope of the present paper. Hence we return to the one-loop renormalization of the $zn$ model.

\paragraph{Renormalization of the FI parameter.}

The first part of the renormalization procedure is similar to the $\CP{N-1}$ model. Indeed, we can extract from the first term the coupling constant renormalization
\beq\label{eq:RenR}
r_{\rm ren}(\mu) = r_0 - \frac{N-\Nt}{2\pi}\log\frac{M}{\mu}\,.
\eeq
The so-called dimensional transmutation occurs at the scale $\Lambda$, when the theory becomes strongly coupled, ($r_{\rm ren}(\Lambda)=0$),
\[
r_0 = \frac{N-\Nt}{2\pi}\log\frac{M}{\Lambda}\,.
\]
Note that this does not happen for $N=\Nt$, the FI parameter remains unchanged and the theory has an IR conformal fixed point.

It was shown in \cite{2010arXiv1004.4049L} that the first Chern class of the $\Nt$-th power of the tautological fiber bundle over $\CP{N-1}$, or in our notation the target space of the HT model, restricted to the base is given by
\[
c_1(M_{\text{HT}})\Big\vert_{\CP{N-1}} = (N-\Nt)\left[\omega_{\CP{N-1}}\right]\,,
\]
where $\left[\omega_{\CP{N-1}}\right]$ denotes the K\"{a}hler class of $\CP{N-1}$. In the above calculations this fact is reflected by \eqref{eq:RenR}. Since in the ${\mathcal N}=(2,2)$ supersymmetric theories 
the K\"{a}hler class is only renormalized at one loop \cite{AlvarezGaume:1985xfa,Friedan:1980jf}, \eqref{eq:RenR} represents the exact answer for the FI term renormalization. Unfortunately one cannot say much 
about the exact part of the K\"{a}hler form. Generally speaking,  it is known to be modified at every order in perturbation theory and its structure is unpredictable unless we carry out an explicit calculation. We will place some argument in the next paragraph about renormalization of such terms at small $|\zeta|^2$.

At this point we can make a
 connection with the $\GLSM$ one-loop computation \eqref{rren}. We have mentioned earlier that in the $\GLSM$ formulation at finite value of the gauge coupling $e$ there are only  two  divergent one-loop graphs which are regularized by the UV cutoff -- the tadpoles emerging from the D-term constraint. The FI renormalization \eqref{rren} was obtained after calculating these tadpoles. 
 Equation \eqref{eq:RenR} confirms this by the corresponding $\NLSM$ calculation performed above. One may now ask if we can trace the origin of the remaining terms in the one-loop $\beta$ function, like the last term in \eqref{eq:logdetznfull}? 

The answer is quite tricky, we will sketch a part of it here. One needs to look more carefully at the perturbation theory at finite $e$. There will be one-loop (and also higher loop) graphs which 
will have $\log ({\mu}/{e})$, where $\mu$ is the IR cutoff (it appears from propagation of light fields in the loops). After we 
make a transition from the $\GLSM$ to the $\NLSM$ by increasing $e$, we will hit the UV cutoff on the way $e\sim M$. 
In $\NLSM$ we identify $M=e$.

 This argument shows us  how additional structures, which were not present in the genuine UV 
 domain of the $\GLSM$ (i.e. the domain above $e$)
 appear in the geometrical renormalization. 
 From the standpoint of the finite-$e$ $\GLSM$ they are of the infrared origin.
 
 Below we will analyze the renormalization of the linear term in $|\zeta|^2$ in \eqref{eq:KahlerPotzn1loop}.

\paragraph{Renormalization of the non-Einstein part.}
 
\begin{figure}[!ht]
\begin{center}
\includegraphics[height=4cm, width=4.5cm]{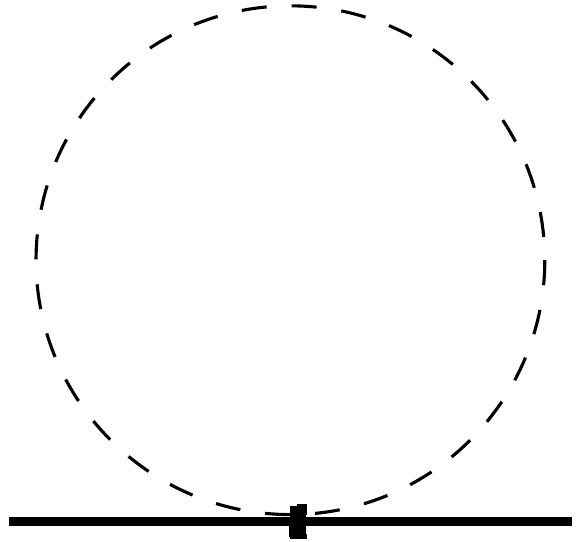}
\caption{Tadpole graphs determining logarithmically divergent contributions to the $\beta$ function
near the origin of the $\CP{N-1}$ space. Two contributions are considered in the text:
(a)   the dashed line represents the $z_j$ fields, while the solid line  $\pt \bar n\pt n$;  and
(b) the dashed line represents the $n_i$ fields, the solid line  $\pt \bar z\pt z$.}
\label{tad}
\end{center}
\end{figure}

Equation \eqref{eq:logdetznfull} gives the exact one-loop answer for the $\beta$ function of the $zn$ sigma model (after applying $\partial_i \bar{\partial}_{\bar{\jmath}}$ to it). Nevertheless it is instructive to understand how the linear term in $|\zeta|^2$ (and higher order terms as well) appear in perturbation theory in geometric formulation. At small $|\zeta|^2$ one can expand the logarithm in the last term in  Eq. \eqref{eq:logdetznfull} to get
\[
-\log\text{det}(g_{i\bar{\jmath}})= (N-\widetilde{N})\log (1+|\Phi_i|^2)-(N-1)|\zeta|^{2}+\mathcal{O}(|\zeta|^4)\,.
\]
Using \eqref{eq:KahlerPotzn1loop} and the coupling renormalization \eqref{eq:RenR} we obtain for the $|\zeta|^2$ term
\[\label{eq:zetasqcontr}
K^{(1)}_{zn}\supset |\zeta|^2 \left(1+\frac{1}{r}\frac{N-1}{2\pi}\log\frac{M}{\mu}\right) = Z|\zeta|^2 \,.
\]
Therefore we can absorb this $Z$ factor by redefining $|\zeta|^2\to |\zeta|^2/Z$.  The contribution \eqref{eq:zetasqcontr} arises in the following calculation.
Since the general structure of the effective action is already known, we can perform a calculation 
at any point in the target space. It is convenient to choose the background field $n_i\to 0$ (while, at 
the same time, $\partial_i n_j\ \neq 0$). Then, as well-known, the logarithmically divergent contribution comes only from the tadpole graphs of the type depicted in Fig.~\ref{tad}.
In the one-loop tadpole graphs the contributions of the second and first terms in (\ref{eq:znKahlerPotential}) in the effective action
are completely untangled from each other. The second term produce just the standard $\CP{N-1}$ renormalization 
of $r$,
\beq
r_{\rm ren}(\mu)= r_0-\frac{N}{2\pi}\,\log {\frac{M}{\mu}}\,,
\label{rrenp}
\eeq
cf. Eq. (\ref{rrenM}). Now, let us examine the impact of the first term in (\ref{eq:znKahlerPotential}). There are two options. We can choose $\partial \bar n \pt n$ as the background and let $z_j$ propagate in the loop (option (a) in Fig.~\ref{tad}) or vice versa. The first option obviously produces
\[
\Delta K^{(1)}_{(a)} = \frac{\tN}{2\pi} \,\log\frac{M}{\mu}\,|\Phi_i|^2\,,
\]
which results in the following term in the renormalized K\"{a}hler potential
\[
\frac{\tN}{2\pi} \,\log\frac{M}{\mu}\, \log\left(1+|\Phi_i|^2\right)\,.
\label{rrenpp}
\]
The difference in signs compared to (\ref{rrenp}) appears from the very beginning. Combining \eqref{rrenp} and \eqref{rrenpp} we recover \eqref{rrenM} or \eqref{eq:RenR}. The second option, with the $n_i$ fields are in the loop, leads us to 
\[
\Delta K^{(1)}_{(b)} = \frac{N-1}{2\pi} \,\log\frac{M}{\mu}\, \sum_{j=1}^{\tN}|\mathfrak{z}_j|^2
\]
which in turn gives
\[
\frac{N-1}{2\pi} \,\log\frac{M}{\mu}\,\, |\zeta |^2\,.
\label{rrenppp}
\]
In the course of  the RG flow from the UV cut-off $M$ down to $\mu$ the first term in the K\"ahler potential (\ref{eq:znKahlerPotential}) acquires the following $Z$ factor
\[
|\zeta|^{2}  \to Z  |\zeta|^{2}  \,,\qquad Z = 1 + \frac{N-1}{2\pi}\log\frac{M}{\mu}\,. 
 \label{mczz}
\]
Thus we can see that at small values of $|\zeta|^2$ the theory can be renormalized at one loop and no counterterm is needed. This is, however, not the case for higher order terms.

\section{Conclusions}\label{Sec:Consclusions}

In this paper we extensively studied the effective theory on semilocal non-Abelian flux tubes in $\ssN=2$ SQCD. We continued the developments of \cite{Shifman:2011xc} where an explicit exact Lagrangian of the corresponding two-dimensional theory ($zn$ model) was derived in a genuinely field theoretic setup. The analysis we have performed in this work for the $zn$ model has been carried 
out  in parallel with the HT model \cite{Hanany:2004ea}. The latter was found  on semilocal vortices 
in a D-brane setup. The bottom line of our investigation is that only the BPS sector  is correctly reproduced by the HT model; the one-loop $\beta$ functions of  $zn$ and HT models coincide. The one-loop $\beta$ function
exhausts the renormalization of the FI term, which means that the exact twisted superpotentials and the BPS spectra of the two models are the same. This result represents the first proof, carried exclusively in a field theory context,
of the correspondence of the
BPS spectra between four dimensional $\ssN=2$ SQCD and the effective theory on the semilocal vortices therein constructed. We also show that the HT and $zn$ model are equivalent in the large-$N (\tN)$ approximation.

The physics beyond the BPS sector is however {\em different}. The difference between the $zn$ and HT models becomes clear when we look at the perturbation theory in the geometric formulation. First of all, the target manifolds of the two models are different, hence their perturbation series do not coincide. We managed to single out a ``corner" in the target 
 space of the two models where the metrics look the same at the leading order in the FI parameter (alternatively, in the vicinity of the origin in the noncompact subspace, see Sec.~\ref{Sec:GeometricFormulations}) for details).
However, far from the origin renormalization coefficients  are completely different. Speaking geometrically, the $zn$ model is a deformation of the HT model in terms of deforming the sections of the bundle (Eqs. \eqref{eq:logdetznfull} and \eqref{eq:HTRicci1loop21} illustrate this).

Contrary to the case of the $\CP{N-1}$ model, where one-loop renormalization can be completely understood in terms of a single coupling renormalization (the  K\"{a}hler class, or the FI term), which is also one-loop-exact, this is not the case both  in the $zn$ and HT models. 
It occurs because both target manifolds are non-Einstein, hence 
the Ricci tensors (which give the one-loop $\beta$ functions) are not proportional to the metric. Nevertheless, due to nice geometric properties of  the fiber bundles (recall that the HT model lives on the total space of the tautological bundle for $\CP{N-1}$), the first Chern class of this bundle is proportional to the K\"{a}hler class with exactly the right coefficient which also appears in the one-loop renormalization of the FI term.

As we discussed in Sec. \ref{Sec:beta} in the $\GLSM$ formulation of both $zn$ and HT models there are only two  divergent graphs (tadpoles), which contribute to the renormalization of the FI parameter \eqref{rrenM}. However, according to the result   \eqref{eq:logdetznfull}, in the $\NLSM$ formulation the one-loop renormalization consists not only of the FI shift, but also from 
the wavefunction renormalization of $\zeta$. Moreover, an additional counterterm is needed in order to fully absorb the one-loop divergence. It is interesting if we relate the two perturbation series in any physically meaningful way. The answer to this question may be negative as, generally speaking, perturbations around a $\GLSM$ fixed point (small gauge coupling) and $\NLSM$ perturbation theory are different. Moreover, the limit $e\to \infty$ leads us away from the perturbative regime of the corresponding $\GLSM$. Still, more detailed perturbative analysis of the gauge theory at finite $e$ is required in order to better understand which Feynman graphs contribute to the UV divergences. This is a suggestive topic for the future research.

\section*{Acknowledgments}
\addcontentsline{toc}{section}{Acknowledgments}
We would like to thank Arkady Vainshtein for numerous fruitful discussions and Kentato Hori for suggesting some useful literature. 
The work of PK, MS,  and WV is supported by the DOE grant DE-FG02-94ER40823.
PK, MS and WV also thank the Galileo Galilei Institute for Theoretical Physics for  hospitality and partial support during 
the Workshop ``Large-$N$ Gauge Theories" where this 
work was completed. PK's work is also supported in part by the Anatoly Larkin Fellowship in Physics at 
the University of Minnesota. The work of AY was  supported by  FTPI, University of Minnesota, by RFBR Grant No. 09-02-00457a and by Russian State Grant for Scientific Schools RSGSS-65751.2010.2.

\appendix
\addcontentsline{toc}{section}{References}
\bibliography{Bibliographysmart}
\bibliographystyle{nb}

\end{document}